\documentclass[iop]{emulateapj}

\slugcomment{{\sc Accepted to ApJ:} July 20, 2012}
\shorttitle{NGC 4649 Abundance Pattern}
\shortauthors{Loewenstein and Davis}

\begin{document}

\title{An In-Depth Study of the Abundance Pattern in the Hot
  Interstellar Medium in NGC 4649}
\author{Michael Loewenstein\altaffilmark{1}} \affil{Department of
  Astronomy, University of Maryland, College Park, MD 20742}
\email{Michael.Loewenstein.1@nasa.gov}

 \and

\author{David S. Davis\altaffilmark{2}} 
\affil{Department of Physics, University of Maryland Baltimore County,
Baltimore, MD 21250}
\email{David.S.Davis@nasa.gov}

\altaffiltext{1}{CRESST and X-ray Astrophysics Laboratory, NASA/GSFC,
Greenbelt, MD.}
\altaffiltext{2}{CRESST and Astroparticle Physics Laboratory,
NASA/GSFC, Greenbelt, MD.}

%==============================================================================

\begin{abstract}
We present our X-ray imaging spectroscopic analysis of data from deep
{\it Suzaku} and {\it XMM-Newton} Observatory exposures of the Virgo
Cluster elliptical galaxy NGC 4649 (M60), focusing on the abundance
pattern in the hot interstellar medium (ISM). All measured elements
show a radial decline in abundance, with the possible exception of
Oxygen. We construct steady state solutions to the chemical evolution
equations that include infall in addition to stellar mass return and
SNIa enrichment, and consider recently published SNIa yields. By
adjusting a single model parameter to obtain a match to the global
abundance pattern in NGC 4649 we infer that introduction of subsolar
metallicity external gas has reduced the overall ISM metallicity and
diluted the effectiveness of SNIa to skew the pattern towards low
$\alpha/Fe$ ratios, and estimate the combination of SNIa rate and
level of dilution. Evidently, newly-introduced gas is heated as it is
integrated into, and interacts with, the hot gas that is already
present. These results indicate a complex flow and enrichment history
for NGC 4649, reflecting the continual evolution of elliptical
galaxies beyond the formation epoch. The heating and circulation of
accreted gas may help reconcile this dynamic history with the mostly
passive evolution of elliptical stellar populations. In an appendix we
examine the effects of the recent updated atomic database {\em AtomDB}
in spectral fitting of thermal plasmas with hot ISM temperatures in
the elliptical galaxy range.
\end{abstract}


%=============================================================================

\section{Introduction}

There are two basic approaches to studying the formation and evolution
of elliptical galaxies. One may directly examine the assembly history
of the baryonic component by observing how the space density and
morphological demographics of the population of elliptical galaxies,
and their progenitors, develop over time. Alternatively one may
conduct detailed investigation of those spectro-photometric properties
that reflect their histories, and the scaling relations among these
properties. While surveys at a range of redshifts indicate significant
growth in both the number and sizes of ellipticals, ``archeological''
investigation finds that elliptical galaxy stellar populations are
mostly in place at high redshifts ($z>2$ for massive systems) and
passively evolve thereafter. Resolution of this apparent paradox is
crucial for validation and elucidation of the prevailing hierarchical
assembly paradigm of galaxy formation, where dynamical evolution and
the triggering of star formation are presumed to be interconnected.

The archeological approach traditionally relies on optical photometry
and spectroscopy of the stellar population; however, the hot
interstellar medium (ISM) provides a rich complementary site of
diagnostic data that is accessible by means of X-ray observation
(Loewenstein \& Davis 2010, hereafter Paper I, and references therein;
Pipino \& Mattecucci 2011, hereafter PM11). The physical properties of
the ISM in giant elliptical galaxies reflect the distinctive history
and nature of these systems and, as such, markedly differ from those
in the ISM of spiral galaxies such as the Milky Way. The most striking
contrast is that, while mass loss from evolved stars is a primary
source of gas in both spirals and ellipticals, most of the mass return
in the latter is promptly heated to the high temperatures
corresponding to the stellar velocity dispersion and does not
currently participate in an ongoing star-gas cycle \citep{mb03}.

This distinct ecology provides a repository of information about
processes in the distant and recent past of elliptical galaxies. The
ISM is more responsive to energetic events than the stellar population
which is, to first order, passively evolving. Evidence of feedback
processes subsequent to the establishment of the stars and the nuclear
supermassive black hole (SMBH) can only be found in the ISM. The ISM
mass within elliptical galaxies is a small fraction of the stellar
mass return integrated over the post-star-formation history of an
elliptical galaxy, implying the existence of some steady and/or
episodic means of gas removal into the intergalactic, or some
circumgalactic, medium. A successful gasdynamical model for
ellipticals must explain ISM that generally are, nevertheless,
sufficiently massive to rule out persistent supersonic or transonic
galactic winds, and also display a large scatter in gas-to-stellar
mass ratio ($M_{\rm gas}/M_{\rm stars}$; Mathews et al. 2006 and
references therein). While ram pressure stripping may be important at
times for some ellipticals in rich clusters, large scale flows driven
by Type Ia supernovae (SNIa) and active galactic nuclei (AGN) likely
dominate the redistribution of the ISM and its removal from the
galaxy.

The amount of energy associated with SNIa exploding at the estimated
rate in ellipticals is sufficient to drive a galactic wind. Although
effects such as depletion through the formation of dust (PM11) may
reduce the discrepancy, the high expected ISM Fe abundance is at odds
with X-ray observations (e.g., Paper I, PM11), and a certain amount of
fine-tuning in the SNIa rate as a function of time is required for
SNIa-driven winds to explain the $M_{\rm gas}/M_{\rm stars}$ scatter
without driving out virtually all of the gas and rendering ellipticals
undetectable as diffuse soft X-ray sources. Feedback from active
galactic nuclei (AGN), implicated in quenching star formation in
ellipticals \citep{sch07,cat09} and establishing the scaling relations
between SMBH mass and stellar mass or velocity dispersion (e.g.,
DeBuhr, Quataert, \& Ma 2011 and references therein), may also drive
galactic flows at later times. AGN interaction with hot ISM is evident
in {\it Chandra} X-ray Observatory observations of several ellipticals
\citep{mn07,nul09}. AGN feedback is fundamentally self-regulating and
intermittent: the SMBH is fueled by an initial inflow that is
subsequently reversed as the AGN powers up, thus cutting off the
supply of gas and enabling the cycle to restart as gas once again
flows inward towards the now-dormant SMBH. There has been great
progress in implementing and applying hydrodynamical simulations that
include various feedback prescriptions \citep{mbb04,bmhb09,cop10},
however it remains to be seen whether the full range of ISM
observables -- X-ray luminosities, temperatures, metallicities and
abundance patterns -- and their galaxy-to-galaxy variations can be
accurately and self-consistently modeled without additional, and
perhaps fundamental, adjustment \citep{mb03}. In addition, many
ellipticals are embedded in an extended hot intergalactic or
circumgalactic medium, and may interact with their environment in a
variety of ways \citep{mj10}. In particular ellipticals may accrete
some combination of primordial gas and gas ejected from winds at
earlier epochs \citep{pm04,dfo11}. Such an external medium may also
confine outflows.

Recent X-ray measurements of elemental abundances in the hot ISM of
ellipticals beg consideration in determining the future direction of
these models, and constraining more general theories of the chemical
evolution of these systems. The ISM abundance pattern may be a
particularly sensitive diagnostic of dynamical processes in
ellipticals. If the sole source of interstellar gas is stellar mass
loss, the ISM abundance pattern in ellipticals will reflect that in
evolved stars found to have $[\alpha/Fe]_{\rm stars}>0$
($[\alpha/Fe]_{\rm stars}$ is defined as the log of the abundance
ratio, with respect to Fe, of elements primarily produced through
$\alpha$ capture to Fe -- relative to the solar ratio). Because the
internal injection and flow of energy, and the various mechanisms of
mass exchange with the external environment, each imprint distinctive
departures from the baseline ISM abundance pattern determined by local
stellar mass loss, this pattern may be analyzed to probe for
signatures of these processes.

Measurement of elliptical galaxy abundance patterns encompassing a
broad range of elements, and extending to large radii may be made with
the {\it Suzaku} Observatory, facilitated by the low internal
background and relatively sharp energy resolution of the XIS CCD
detectors. In Paper I we derived the abundance pattern in the
elliptical galaxy NGC 4472 from analysis of {\it Suzaku} spectra,
supported by analysis of co-spatial {\it XMM-Newton} Observatory
spectra. Application of simple chemical evolution models to these
data, led us to conclude that the abundances may be explained by a
combination of $\alpha$-element enhanced stellar mass loss and direct
injection of ejecta from SNIa exploding at a rate $\sim 4-6$ times
lower than the standard value. In addition, we discovered abundance
anomalies in the sense that no published set of SNIa yields could
simultaneously reproduce the inferred Ca and Ar, and Ni abundances;
and (confirming prior results summarized in Paper I) in the sense that
standard core collapse nucleosynthesis models evidently overproduce O
by $\sim 2$.

In this paper we adopt a broadly similar approach in our investigation
of NGC 4649 (M60). As is the case for NGC 4472, NGC 4649 is a giant
elliptical galaxy in the Virgo cluster with an old stellar population
enhanced in $\alpha$-elements -- but with several notable differences
both optically and in X-rays. In particular NGC 4649 has a substantial
major axis rotation (Brighenti et al. 2009, and references therein),
and is considerably more compact in X-rays. We have adjusted our
spectral analysis procedures in response to the distinctive X-ray
characteristics of NGC 4649, and have revised and expanded our models
to consider possible episodes of inflow, and a recently published set
of SNIa yields.

We detail our data reduction and spectral analysis procedures in
Sections 2 and 3, where we present our derived NGC 4649 hot ISM
thermal and chemical properties and their radial variation. In Section
4 we focus on interpreting the global abundance pattern in the context
of the relative contributions of metal enrichment from stellar mass
return, SNIa, and inflow of extragalactic material using steady state
solutions to the equations of chemical evolution. Section 5 includes a
summary of our conclusions, and discusses possible implications of our
results. In an appendix we examine the effects of the recent updated
atomic database {\em AtomDB} in spectral fitting of elliptical galaxy
hot ISM such as NGC 4649.

\section{Construction of NGC 4649 Spectra and Associated Files}

\subsection{{\it Suzaku} Spectral Extraction and Preparation}

NGC 4649 was observed with {\it Suzaku} \citep{mit07} between
2006-12-29 and 2007-01-04 (ObsID 801065010), at which time three
co-aligned, $17.8'\times 17.8'$ field-of-view X-ray Imaging
Spectrometer (XIS) CCD cameras \citep{koy07} -- two front-illuminated
(FI: XIS0 and XIS3) and one back-illuminated (BI: XIS1) -- were
operational, each XIS in the focal plane of an X-ray Telescope (XRT)
with a $2'$ half-power diameter \citep{ser07}. Observations were
conducted utilizing the space-row charge injection (SCI) technique
that reverses the degradation in energy resolution caused by
accumulated radiation damage \citep{nak08}. We initiate our data
reduction with the unfiltered event files. These data underwent
Version 2.0.6.13 pipeline processing on 2007-08-17 that enables one to
properly account for the effect of SCI on instrument characteristics
and performance \citep{uch08}. We reprocess the unfiltered event files
by hand in order to apply updated calibration data and software,
following the procedures described in Paper I. We recalculate PI
values and grades, select event grades (0, 2, 3, 4, 6) that correspond
to X-ray photon events, filter on pixel status (eliminating bad charge
transfer efficiency columns, and rows invalidated by the charge
injection process) and select good time intervals (GTI) based on
pointing, data and telemetry rates, SAA proximity (``$\rm{SAA}\_ {\rm
HXD}\equiv 0, \rm{T}\_ \rm{SAA}\_ \rm{HXD}> 436$''), and proximity to
the earth's limb and illuminated Earth (``$\rm{ELV}> 5, \rm{DYE}\_
\rm{ELV}> 20$''). In addition, telemetry-saturated frames and
calibration source photons are screened out; and, hot and flickering
pixels are removed, and we accept only GTI where the revised
geomagnetic cut-off rigidity COR2 $>4$, thus eliminating intervals
with the highest particle background level \citep{tawa08} without
compromising overall statistical accuracy. $5\times 5$ event files are
converted to $3\times 3$ mode format, and merged with the $3\times 3$
event files. The exposure times in the cleaned event files are 216.6
(216.5) ks for the XIS0 and XIS3 (XIS1) spectra.

We extract spectra from the inner $8'$ ($\sim 6.5R_e=36.2$ kpc at the
NGC 4649 distance of 15.6 Mpc; Tonry et al. 2001, where $R_e$ is the
optical half-light radius from Bregman, Temi, \& Bregman 2006) in $2'$
and $4'$ circular annuli, centered on the NGC 4649 optical nucleus
($\alpha=12h43m40.0s$, $\delta=+11^{\circ}33'10''$) that very closely
corresponds to the X-ray peak in the {\it Suzaku} image. The spectral
redistribution matrix files ({\bf rmf}) are generated using {\tt
xisrmfgen} version 2009-02-28. The {\bf rmf} and spectral files are
binned to 2048 channels. The effective area function files ({\bf arf})
for the source spectra are generated by the {\tt xissimarfgen} version
2009-01-08 Monte Carlo ray-tracing program \citep{ish07} with 400000
simulation photons per energy bin, and an input source fits image file
generated from a $\beta$-model fit to the background-subtracted X-ray
surface brightness profile extracted from archival {\it Chandra} data
($\beta=0.467$, core radius $3.5''$). Spectra from the FI chips, XIS0
and XIS3, are co-added and a weighted XIS0+3 response function
calculated from their respective {\bf rmf} and {\bf arf} files. Source
spectra are grouped into bins with a minimum of 15 cts, and we derive
best-fit parameters and confidence levels using $\chi^2$ statistics.
Errors correspond to 90\% confidence limits.

The XIS background consists of Non-X-ray (instrumental and charged
particle) Background (NXB), Galactic X-ray Background (GXB), and
(extragalactic) Cosmic X-ray Background (CXB) components. Since NGC
4649 fills the {\it Suzaku} field of view, we estimate and subtract
the NXB component and include additional components in our spectral
fits to account for the GXB and CXB. The NXB component is estimated
from observations of the night earth taken in SCI mode within 150 days
of the starting or ending dates of our observation using {\tt
  xisnxbgen} version 2008-03-08. The NXB event list in that time
interval undergoes the identical screening as the source data, is
sorted by geomagnetic cut-off rigidity, and weighted according to the
cut-off rigidity distribution in the source event file
\citep{tawa08}. The estimated NXB spectra include only those events
collected in the regions on the detector from which the source spectra
are extracted. Since the CXB and GXB are spatially distributed
distinctly from that of the NGC 4649 X-ray emission, a separate {\bf
  arf} file is generated using 2000000 simulation photons per energy
bin from a uniform source of radius $20'$ and applied to the
background in spectral fits. Combined XIS0+3 NXB spectra and response
functions are constructed as described above.

We display the NXB-subtracted source count rates in each annulus, and
the fraction of the total counts in the NXB for the corresponding
bandpass used in spectral fitting (see below), in Table 1 .

\begin{deluxetable}{lll}
\tabletypesize{\scriptsize}
\tablewidth{0pt}
\tablecaption{{\it Suzaku} Count Rates}
\tablehead{\colhead{region} & \colhead{XIS0+3} & \colhead{XIS1}}
\startdata 
$0-2'$ & 0.399  (0.008) & 0.299  (0.011) \\
$2-4'$ & 0.154  (0.058) & 0.127  (0.069) \\
$4-6'$ & 0.0840 (0.18)  & 0.0676 (0.21)  \\
$6-8'$ & 0.0583 (0.18)  & 0.0522 (0.22)  \\
$4-8'$ & 0.151  (0.22)  & 0.124  (0.25)  \\
$2-6'$ & 0.232  (0.10)  & 0.189  (0.125) \\
\enddata 
\tablecomments{Total counts rates in cts s$^{-1}$ in the energy band
  used for spectral fits (0.4-4.0 keV for the $6-8'$ annulus, 0.4-7.0
  keV for all others). Values in parentheses represent the fractions
  of total counts in the NXB.}
\end{deluxetable}

\subsection{{\it XMM-Newton} EPIC Spectral Extraction and Preparation}

NGC~4649 was observed with {\it XMM-Newton} \citep{jan01} for $\sim$54
ks on 2001-01-02 (ObsID: 0021540201), and for $\sim$92 ks on
2007-12-19 (ObsID: 0502160101). We downloaded the data from the High
Energy Science Archive Research Center ({\tt HEASARC}) data
archive\footnote{http://heasarc.gsfc.nasa.gov/docs/archive.html} and
processed them with Science Analysis System (SAS) version 11.0.1,
utilizing the techniques described in \cite{ks08} and \cite{sno08}
through application of the {\it XMM-ESAS} (XMM-Newton Extended Source
Analysis Software) suite of procedures. EPIC-MOS and -PN CCD camera
photon event files were generated in the standard way with the tasks
{\tt emchain} and {\tt epchain},
respectively\footnote{http://xmm.esac.esa.int/external/xmm\_user\_support/\\documentation/sas\_usg/USG/}. Files
were processed with $FLAG == 0$, and screened to retain only those
events with $PATTERN<=12$ for the MOS and $PATTERN<=4$ for the PN.
 
The event files for each observation was screened for flares by
examining their lightcurves using the {\tt xmmlight\_clean} scripts
\citep{bpt08}. The useful exposure times, after cleaning, are $\sim$49
ks (MOS) and $\sim$41 ks (PN) for the ObsID 0021540201 event lists,
and $\sim$73 ks (MOS) and $\sim$70 ks (PN) for the ObsID 0502160101
event lists. {\it XMM-Newton} observations may be significantly
contaminated by soft protons originating in the solar wind
\citep{ks08}. Although the lightcurve filtering process largely
removes soft proton flares, we determine the level of residual soft
proton contamination using the {\tt Fin\_over\_Fout} method
\citep{mdl04}. We find the level of contamination to be minor for the
MOS detectors, and negligible for the PN detector, during the first
(2001) observation. However for the second (2007) observation,
although the two MOS cameras again experienced only minor
contamination, the PN camera was significantly affected.

The extracted spectra are binned so that each channel has a minimum of
50 counts, and we consider an energy range of 0.3-5.5 keV for the MOS
detectors and 0.4-5.5 keV for the PN detector.  We extracted spectra
from concentric annuli with identical centers and (2${^\prime}$)
widths to those of the {\it Suzaku} extraction regions.  The spectral
extraction was performed using the {\it XMM-ESAS} tasks {\it
  mos-spectra} and {\it pn-spectra} for the relevant detectors. These
scripts extract the spectra, and generate the instrument response
({\bf rmf} and {\bf arf}) matricies. The tasks {\it mos-back} and {\it
  pn-back} are used to model the instrumental and particle backgrounds
(NXB) for each of the extracted spectra.

The source count rates in each annulus are displayed in Table 2.

\begin{deluxetable}{lllllll}
\tabletypesize{\scriptsize}
\tablewidth{0pt}
\tablecaption{{\it XMM-Newton} EPIC Count Rates}
\tablehead{\colhead{region} & \colhead{MOS1} & \colhead{Exposure}&
  \colhead{MOS2} & \colhead{Exposure}&\colhead{PN}&
  \colhead{Exposure}}
\startdata 
\multicolumn{7}{c}{ObsID 0021540201} \\
$0-2'$ & 0.518 & 45.21 & 0.507 & 45.54  & 1.797 & 49.28 \\
$2-4'$ & 0.173 & \nodata & 0.185 & \nodata & 0.672 & \nodata \\
$4-6'$ & 0.103 & \nodata & 0.103 & \nodata & 0.192 & \nodata \\
$6-8'$ & 0.067 & \nodata & 0.059 & \nodata & 0.313 & \nodata \\\hline
\multicolumn{7}{c}{ObsID 0502160101} \\ 
$0-2'$ & 0.504 & 67.24 & 0.482 & 66.46 & 1.782 & 58.25 \\
$2-4'$ & 0.154 & \nodata & 0.142 & \nodata & 0.447 & \\
$4-6'$ & 0.103 & \nodata & 0.080 & \nodata & 0.201 & \nodata \\
$6-8'$ & 0.056 & \nodata & 0.057 & \nodata & 0.128 & \nodata \\
\enddata 
\tablecomments{Total counts rates in cts s$^{-1}$ in the energy band
  used for spectral fits (0.4-5.5 keV for the PN and
  0.3-5.5 keV for the MOS). The exposure time is in units
  of ks.}
\end{deluxetable}

We use the the {\tt HEASARC} background
tool\footnote{http://heasarc.gsfc.nasa.gov/cgi-bin/Tools/xraybg/xraybg.pl}
to extract the ROSAT All-Sky Survey (RASS) spectrum from an annular
region, with an inner radius of 1$\arcdeg$ and an outer radius of
2$\arcdeg$, centered on the X-ray peak of NGC~4649 -- assuming that
the average in the annulus is a fair representation of the CXB and GXB
at the position of the galaxy. This data is simultaneously fit with
the EPIC data.

\subsection{{\it XMM-Newton} RGS Spectral Extraction }

We extract the RGS spectra for this galaxy using the standard SAS
procedures, applying the {\tt rgsproc} task to generate filtered event
files and spectra for both observations (OBSID 0021540201 and
0502160101). We exclude some short time intervals where we detected
flares in the background light curves, resulting in the exposure times
listed in Table~3.

We use the standard background products and generate RGS response
matrices using the task {\it rgsrmfgen} that are technically
applicable to spectra of point sources in the RGS field of view. To
account for the mixing of spatial and spectral information from
extended sources, the matrices are convolved with an image of NGC~4649
(with point-sources removed) created from {\it Chandra} ACIS-S data
(OBSID: 8182) using the {\tt FTOOL} {\it rgsrmfsmooth}. The image is
adaptively smoothed with minimum of 75 counts to reduce the noise
prior to the response matrix convolutions.

\begin{deluxetable}{llllll}
\tabletypesize{\scriptsize}
\tablewidth{0pt}
\tablecaption{{\it XMM-Newton} RGS Count Rates}
\tablehead{\colhead{region} & \colhead{RGS1} & \colhead{Exposure}&
  \colhead{RGS2} & \colhead{Exposure}&\colhead{Order}}
\startdata 
\multicolumn{6}{c}{ObsID 0021540201} \\
$0-1\farcm 5$ & 0.064 & 52.56 & 0.100 & 51.08  & 1 \\
$0-1\farcm 5$ & 0.026 & \nodata & 0.028 & \nodata  & 2 \\
\multicolumn{6}{c}{ObsID 0502160101} \\ 
$0-1\farcm 5$ & 0.051 & 74.73 & 0.090 & 74.78 & 1 \\
$0-1\farcm 5$ & 0.026 & \nodata & 0.027 & \nodata & 2 \\
\enddata 
\tablecomments{Total counts rates in cts s$^{-1}$ for the RGS spectra
for each observation. The exposure time is in units of ks.}
\end{deluxetable}

\section{Spectral Analysis}

\subsection{\it Suzaku Spectral Analysis}

We employ {\sc Xspec} version 12.7 to simultaneously fit the {\it
  Suzaku}~XIS0+3 and XIS1 NGC 4649 spectra. Our baseline source model
consists of an absorbed single-temperature thermal plasma, plus a 7.5
keV thermal bremsstrahlung component to account for the LMXB
\citep{irw03,hb08}. Temperatures and heavy elemental abundances (C, N,
O, Ne, Mg, Al, Si, S, Ar, Ca, Fe, Ni) in the former component are
estimated using the {\bf vapec} model that, in this latest version of
{\sc Xspec}, incorporates the updated {\em AtomDB} v2.0.1 atomic
database. We note that, at elliptical galaxy X-ray temperatures,
spectral fits are significantly impacted by changes with respect to
the {\em AtomDB} v1.3.2 database used in earlier versions of {\sc
  Xspec} (see Appendix A). Therefore, any detailed comparison with
previous results \citep{rsi04,rsi06,tf08,hum08,ds08,nm09} are of
limited utility, although we note a general $<4'$ ISM temperature
profile consistency if we use {\em AtomDB} 1.3.2. Addition of a second
{\bf vapec} model (with identical abundances) does not improve the
fits, and is not further considered. We fix the redshifts of the
source components at the value corresponding to the NGC 4649
distance. We apply the \cite{wam00} Tuebingen-Boulder ISM absorption
model ({\bf tbabs}), and values in excess of the Galactic value of
$2.2\times 10^{20}$ cm$^{-2}$ \citep{dl90} are considered as described
below. We adopt the \cite{agss09} solar abundance standard, newly
implemented in {\sc Xspec 12.6}, fixing C and N at their solar
values. Since we find that only O, Ne, Mg, Si, S, and Fe are
(formally) well-constrained, Al:Mg, Ca:Ar:S, and Ni:Fe are fixed at
their solar ratios \citep{mur11}. The models include a constant
XIS1/XIS03 multiplicative factor to account for flux calibration
offsets that, when fitted for, ranges from 1.0--1.08.

We include four astrophysical background components for all extraction
regions, although these are not always formally required for the
innermost annuli where the contribution of one or more of these is
small or negligible. We model the CXB by an index 1.4 power-law
\citep{kush02}, and the GXB by the sum of solar-abundance thermal
plasma ({\bf apec}) models representing the Milky Way halo (MWH) and
Local Hot Bubble (LHB) \citep{hay09,nm10}. The CXB and MWH are
absorbed by the Galactic column, the LHB is unabsorbed. We also
include an additional extended, 0.5 solar abundance {\bf apec}
component \citep{lvm01} to model the Virgo ICM (see, also Buote \&
Fabian 1998, Randall et al. 2006, 2008, Ji et al. 2009 -- hereafter
Ji09, and Bogd\'an \& Gilfanov 2011, for evidence of emission from a
second, hotter medium). We allow normalizations for these components
in each annuli to freely vary in general; however, the MWH/LHB flux
ratio is assumed to be constant (see below). We tie the background
XIS1-to-XIS0+3 ratio to the value in the source model. Separate {\bf
  arfs} are applied to background and source model components as
described above.

Because the GXB is most prominent in the outermost annuli, we fit the
$6-8'$ annulus (simultaneously with the RASS spectrum) first to
determine the LHB, MWH, and ICM temperatures. The NXB dominates above
4 keV in this annulus, so we fit over the 0.4-4 keV bandpass (all
other spectra are fit over 0.4-7 keV). The column density is fixed at
the Galactic value, however {\it post facto} fits with the higher
absorption determined from the innermost annulus (see below) yield
virtually identical results. The background-component temperatures in
the spectra extracted from other regions are fixed at these outer
region best-fit values: $kT({\rm LHB})=0.105$, $kT({\rm MWH})=0.187$,
$kT({\rm ICM})=1.69$, as is the MWH/LHB flux ratio
\citep{mur11}. Relaxing these constraints did not significantly impact
the best-fit ISM parameters.

\begin{deluxetable*}{cccccccccc}
\tabletypesize{\scriptsize}
\tablewidth{0pt}
\tablecaption{{\it Suzaku} Best-Fit ISM Parameters}
\tablehead{\colhead{region} & \colhead{$\chi^2/{\nu}$} &
  \colhead{$N_H$} & \colhead{$kT$} & \colhead{0} & \colhead{Ne} &
  \colhead{Mg} & \colhead{Si} & \colhead{S} & \colhead{Fe}}
\startdata

{$0-2'$} & 1994/1624 & $8.4_{-0.6}^{+0.3}$ & $0.867_{-0.003}^{+0.003}$
& $0.91_{-0.12}^{+0.09}$ & $2.09_{-0.11}^{+0.19}$ &
$1.14_{-0.04}^{+0.05}$ & $1.01_{-0.034}^{+0.05}$ &
$1.20_{-0.10}^{+0.15}$ & $0.99_{-0.01}^{+0.02}$\\

{$2-4'$} & 1578/1463 & $10.0_{-2.2}^{+2.3}$ &
$0.890_{-0.016}^{+0.020}$ & $0.76_{-0.17}^{+0.17}$ &
$1.24_{-0.18}^{+0.24}$ & $0.49_{-0.07}^{+0.06}$ &
$0.57_{-0.06}^{+0.06}$ & $0.76_{-0.16}^{+0.16}$ &
$0.47_{-0.04}^{+0.08}$\\

{$4-6'$} & 1529/1449 & $2.2*$ & $0.931_{-0.047}^{+0.037}$ &
$0.91_{-0.31}^{+0.42}$ & $0.72_{-0.50}^{+0.37}$ &
$0.36_{-0.15}^{+0.20}$ & $0.62_{-0.15}^{+0.18}$ &
$0.56_{-0.43}^{+0.46}$ & $0.34_{-0.10}^{+0.06}$\\

{$6-8'$} & 1054/1107 & $2.2*$ & $0.871_{-0.031}^{+0.083}$ &
$0.62_{-0.46}^{+0.33}$ & $0.046_{-0.046}^{+0.33}$ &
$0.21_{-0.16}^{+0.15}$ & $0.34_{-0.17}^{+0.16}$ &
$0.38_{-0.38}^{+0.57}$ & $0.15_{-0.02}^{+0.09}$\\

{$2-6'$} & 2040/1946 & $9.7_{-2.7}^{+2.6}$ & $0.897_{-0.018}^{+0.019}$
& $0.84_{-0.21}^{+0.16}$ & $1.01_{-0.16}^{+0.23}$ &
$0.44_{-0.07}^{+0.05}$ & $0.54_{-0.06}^{+0.05}$ &
$0.68_{-0.15}^{+0.15}$ & $0.41_{-0.03}^{+0.06}$\\

{$4-8'$} & 2019/2018 & $2.2*$ & $0.923_{-0.062}^{+0.025}$ &
$0.68_{-0.19}^{+0.32}$ & $0.31_{-0.22}^{+0.17}$ & $0.28_{-0.09}^{+0.14}$
& $0.44_{-0.09}^{+0.17}$ & $0.43_{-0.29}^{+0.34}$ &
$0.22_{-0.023}^{+0.02}$\\

\enddata \tablecomments{Displayed for each spectral extraction region
are minimized $\chi^2$/dof for simultaneous fits to {\it Suzaku} XIS0,
XIS3, and XIS1 spectra, best-fit columns densities (in units of
$10^{20}$ cm$^{-2}$), best-fit ISM temperatures (in keV), and best-fit
O, Ne, Mg, Si, S, and Fe abundances referred to the solar standard of
\cite{agss09}. Asterisks denote fixed parameters.}
\end{deluxetable*}

\begin{figure}[ht]
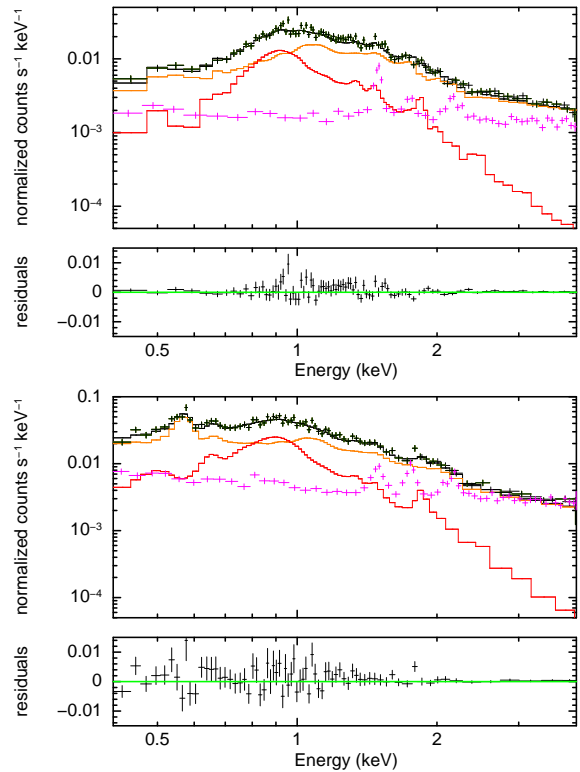

\centering
\includegraphics[scale=0.33,angle=-90]{n4649_xis03_6to8_spec.ps}
\hfil
\includegraphics[scale=0.33,angle=-90]{n4649_xis1_6to8_spec.ps}
\caption{{$6-8'$ aperture {\it Suzaku} XIS source
    (NXB-subtracted) spectra and best simultaneous fit models to data
    from all three detectors (black data points and histogram in top
    panel), with residuals (bottom panel). Contributions from the
    ensemble of low mass X-ray binaries (if significant; LMXB, blue),
    from hot ISM thermal plasma components (red), and from the sum of
    GXB, CXB, and ICM background components (orange) are separately
    plotted. The NXB (purple), that is subtracted prior to spectral
    fitting, is also displayed. The {\it top} panel (a) shows the sum
    of the XIS0 and XIS3 spectra, folded through a weighted response,
    the bottom panel (b) shows the XIS1 spectrum.}}
\end{figure}

\begin{figure}
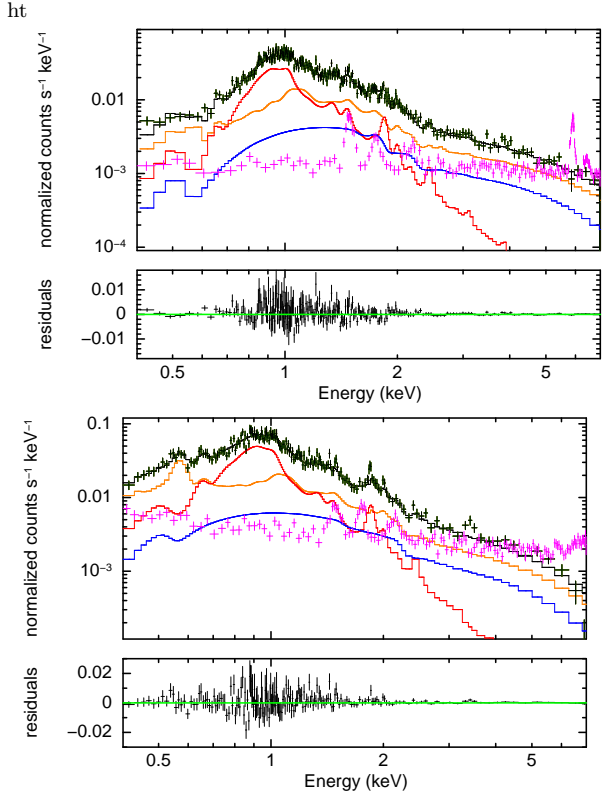
{ht}
\centering
\includegraphics[scale=0.33,angle=-90]{n4649_xis03_4to6_spec.ps} 
\hfil
\includegraphics[scale=0.33,angle=-90]{n4649_xis1_4to6_spec.ps}
\caption{{Same as Figure 1 for the $4-6'$ aperture.}}
\end{figure}

\begin{figure}[ht]
\centering
\includegraphics[scale=0.33,angle=-90]{n4649_xis03_2to4_spec.ps}
\hfil
\includegraphics[scale=0.33,angle=-90]{n4649_xis1_2to4_spec.ps}
\caption{{Same as Figure 1 for the $2-4'$ aperture.}}
\end{figure}

\begin{figure}[ht]
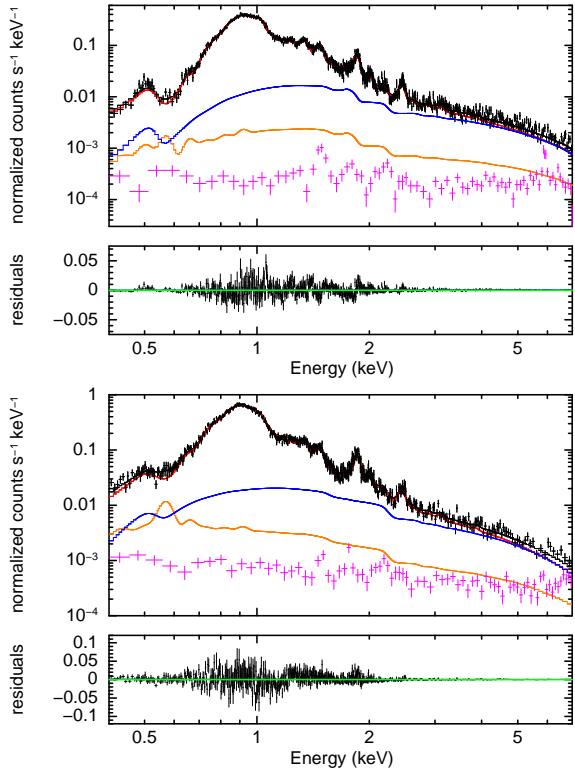

\centering
\includegraphics[scale=0.33,angle=-90]{n4649_xis03_0to2_spec.ps}
\hfil
\includegraphics[scale=0.33,angle=-90]{n4649_xis1_0to2_spec.ps}
\caption{{Same as Figure 1 for the $0-2'$ aperture.}}
\end{figure}

We display the minimized (reduced) $\chi^2$ values, and source model
temperatures abundances, and column densities, for each annuli in
Table 4; and plot spectra, best-fit models (and their components), and
data residuals to the best-fit model for the $2'$ annular partitioning
of the $8'$ region in Figures 1-4. We note the following details. We
find degeneracies in some fits among the normalizations of the CXB,
ICM, and LMXB hard component normalizations (thus the latter component
does not appear in Figures 1 and 3); however, these do not introduce
significant errors in the hot ISM abundance or temperature
determinations. Best-fit model parameters of interest are insensitive
to whether thermal bremsstrahlung or power-law models are adopted for
LMXB components. We adopt the former since in this case fits are
somewhat less sensitive to assumptions about the absorption. In the
$0-2'$ region, we found systematic positive residuals at $\sim$
1.23-1.24 keV that can be well fit by a narrow ({\it i.e.} unresolved
by {\it Suzaku}) Gaussian emission line. We include this component in
the fits plotted and reported here. This excess may be a sign of
lingering inadequacies in calibration (via the instrumental response
or background) or in the plasma code; and, modeling it only effects
the quality of fit (a reduction of $\chi^2$ by $\sim 100$) and does
not alter the best-fit parameters. Excess absorption is strongly
required (only) in the inner $4'$ -- $\Delta \chi^2\sim 260$ (26) in
the $0-2'$ ($2-4'$) region. The radial variation argues against the
cause being incorrect modeling of the contamination of the
time-dependent Optical Blocking Filter, which we further confirmed by
fitting the $0-2'$ spectra extracted from data reprocessed using a
recently updated (2011 June) contaminant model. We have also applied
the mixing model in {\sc Xspec} ({\tt suzpsf}) that aims to account
for the cross-contamination of spectra due to the broad {\it Suzaku}
point spread function (psf). These solutions tend to be unstable in
how the ISM abundances are distributed; however, we note that the
emission-averaged abundances over the entire ($0-8'$) region are
robust and consistent with those derived from the above fits.

We plot the {\it Suzaku}-derived abundance profiles of O, Ne, Mg, Si,
S, and Fe in Figures 5abc, and abundance ratio profiles (with respect
to Fe) in Figures 6ab. Since these profiles are not deprojected, nor
corrected for the {\it Suzaku} point spread function, the true
abundance gradients are steeper. There is clearly a negative abundance
gradient for Ne, Mg, Si, S, and Fe, and a hint of positive gradients
in S/Fe and Si/Fe (Figure 6b).

\begin{figure}[ht]
\centering
\includegraphics[scale=0.33,angle=0]{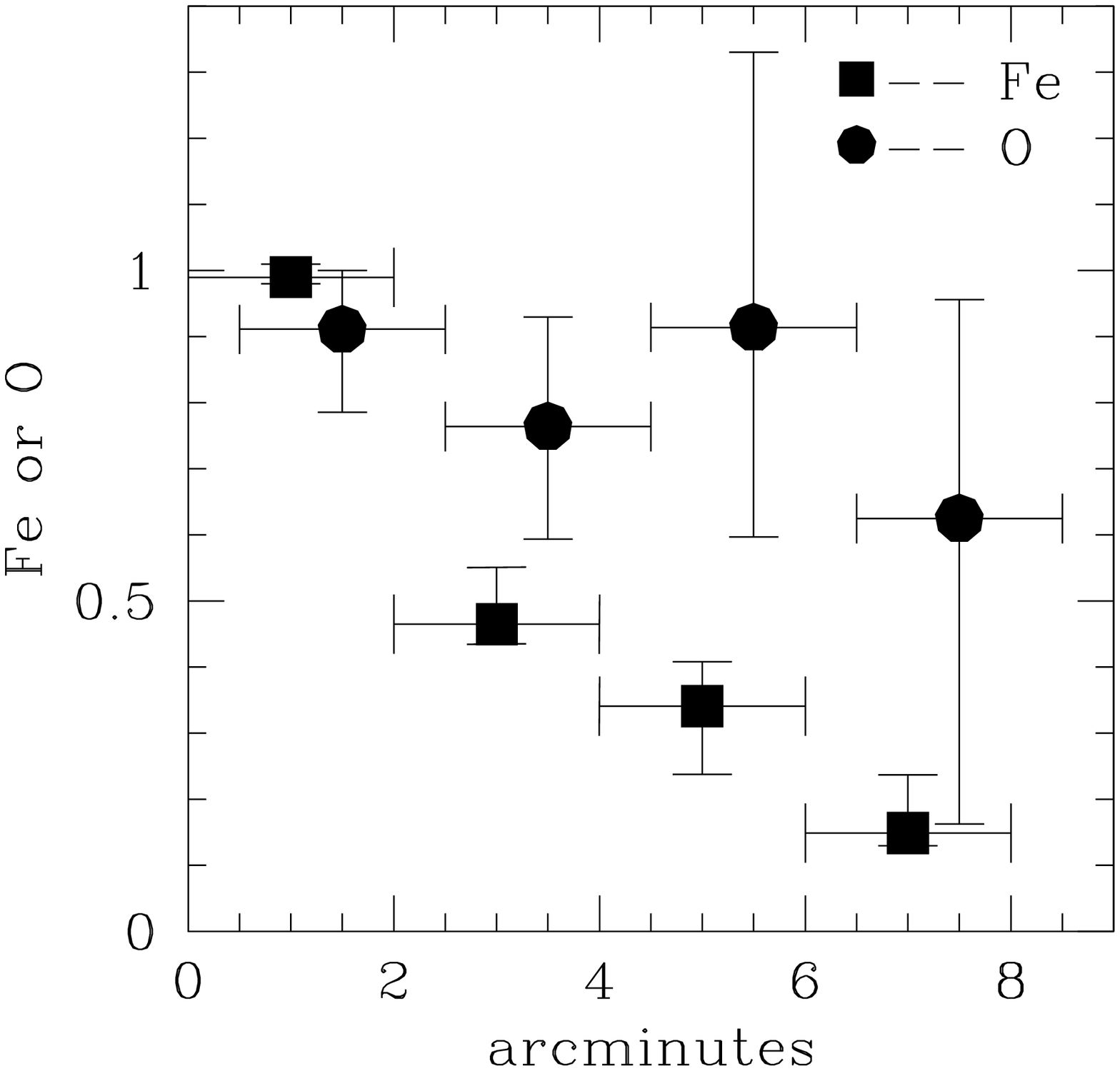}\hfil
\includegraphics[scale=0.33,angle=0]{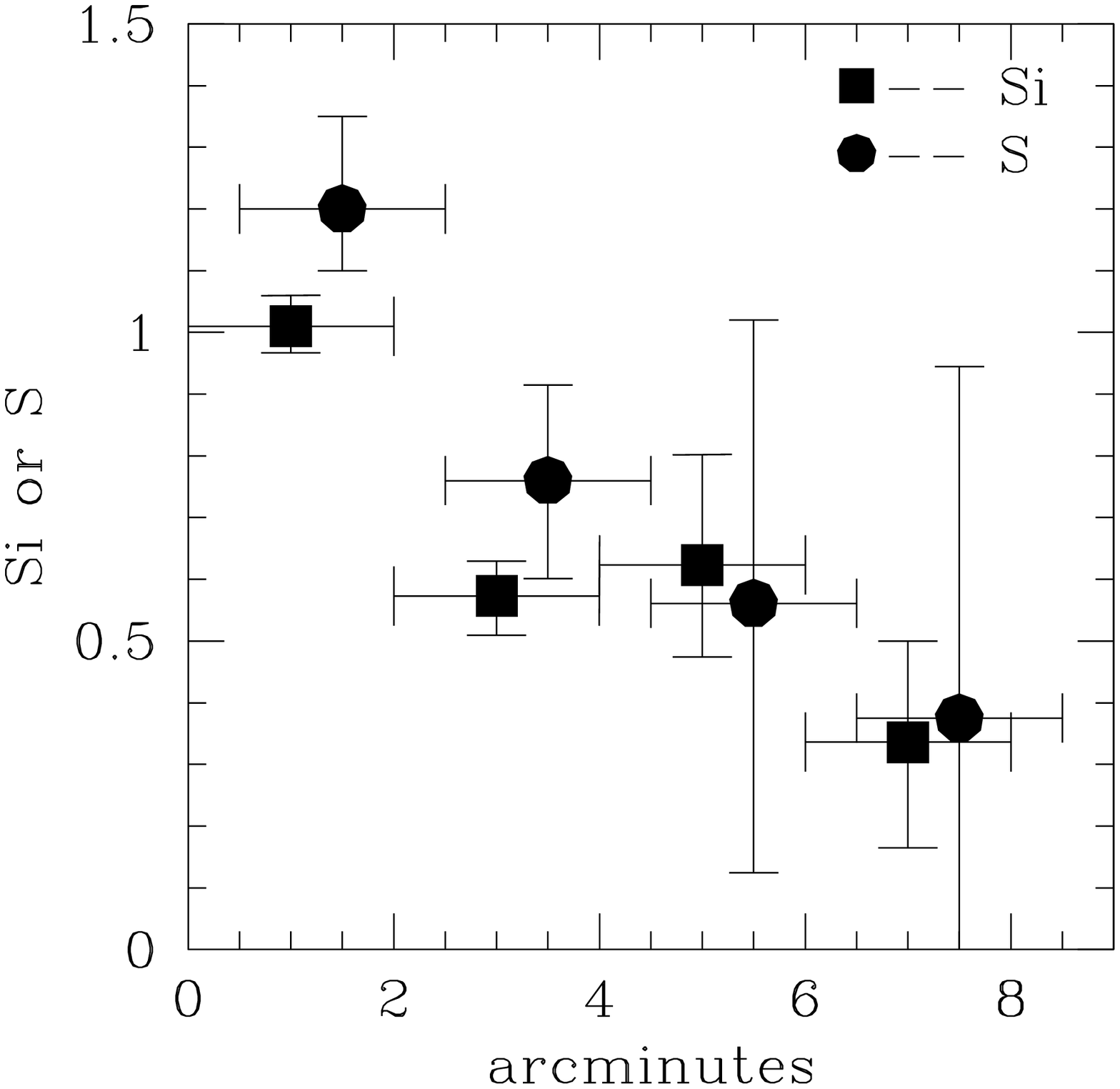}\hfil
\includegraphics[scale=0.33,angle=0]{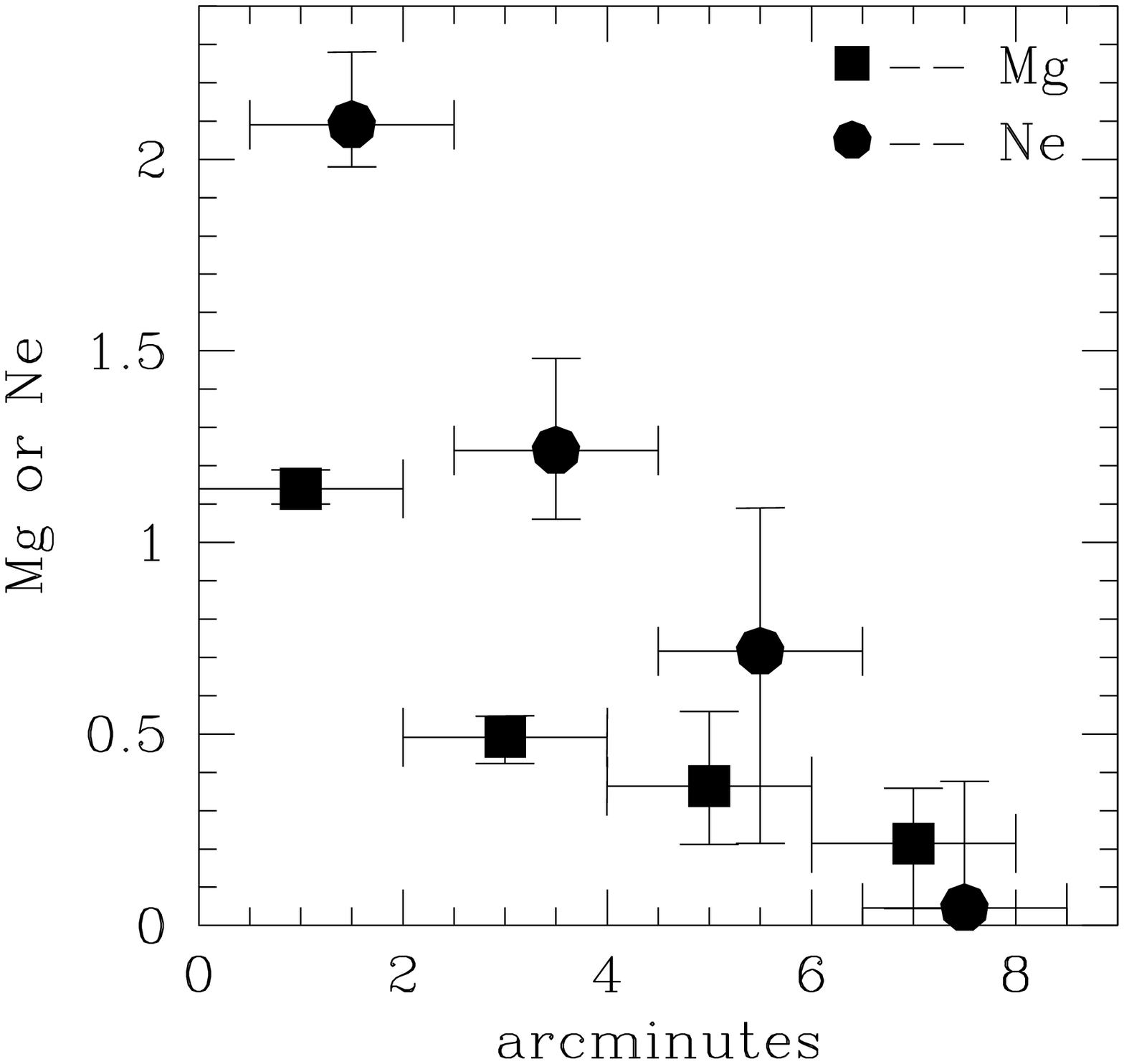}
\caption{{Fe and O ({\it top} panel, (a)), Si and S
    ({\it middle} panel, (b)), Mg and Ne ({\it bottom} panel, (c))
    abundance profiles. The points represented by circles are shifted
    to the right by $0.5'$.}}
\end{figure}

\begin{figure}[ht]
\centering
\includegraphics[scale=0.4,angle=0]{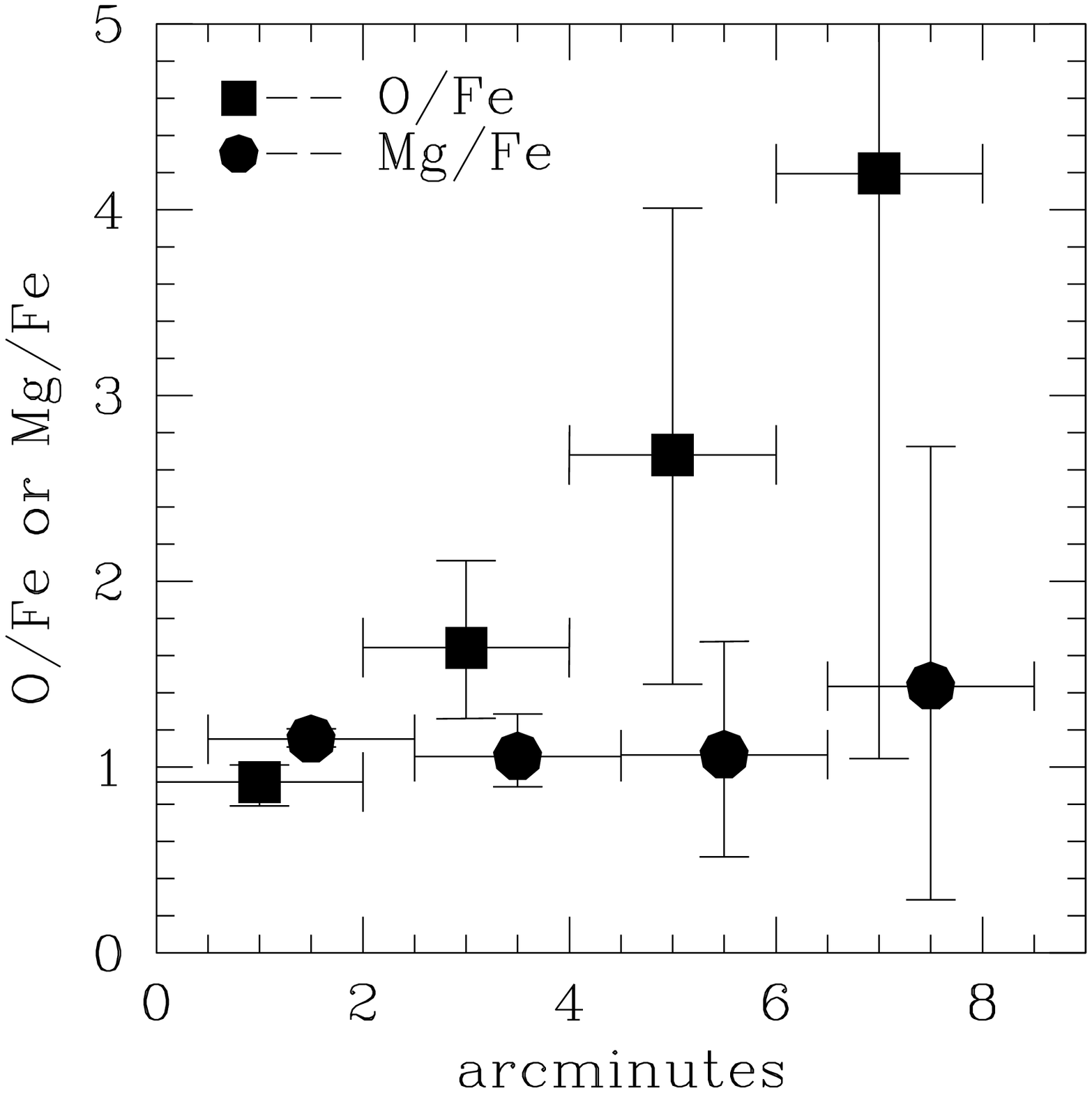}\hfil
\includegraphics[scale=0.4,angle=0]{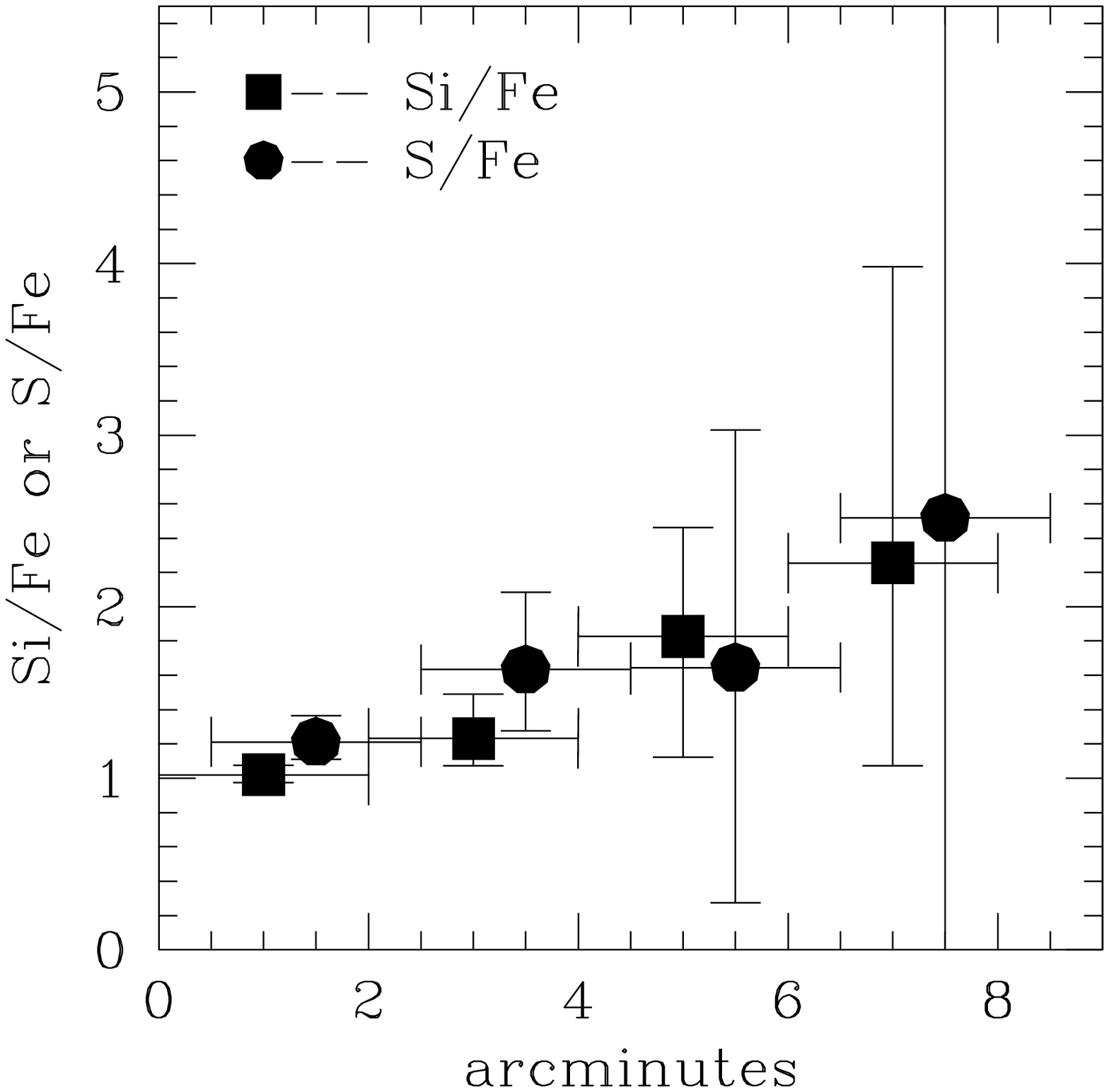}
\caption{{Profiles of abundance ratios with respect to
    Fe for O and Mg ({\it top} panel, (a)); and, Si and S ({\it
      bottom} panel, (b).}}
\end{figure}

We plot the {\it Suzaku}-derived abundance ratios with respect to Fe
for the entire galaxy (out to $8'$) in Figure 7, estimated as an
emission-measure-weighted average of the separate $0-2'$, $2-4'$,
$4-6'$, and $4-8'$ spectra (consistent results are derived using the
$0-2'$, $2-4'$, and $4-8'$ regions, or the $0-2'$, $2-6'$, and $6-8'$
regions). In subsequent sections we use this as a proxy for the
gas-density-averaged abundance. Analytic and semi-analytic estimates
utilizing the $\beta$-model fit to the surface density (see above),
and a range of abundance gradients consistent with the data for
various elements, show that the emission-measure approximation leads
to overestimates of less than 25\%. For comparison, the NGC 4472
pattern (renormalized to the same solar standard, but averaged over
$\sim 4.6R_e$) from Paper I is also shown. Although the average Fe
abundance is 0.51 for NGC 4649 and 1.65 for NGC 4472, the Mg/Fe,
Si/Fe, and S/Fe abundance ratios are consistent between the two
galaxies. However, the agreement is poorer for the more problematic
O/Fe and Ne/Fe ratios (O/Fe because of the importance of Galactic halo
emission in the region of the O lines, Ne/Fe because of blending of Ne
with Fe L lines).

\begin{figure}[ht]
\centering
\includegraphics[scale=0.4,angle=0]{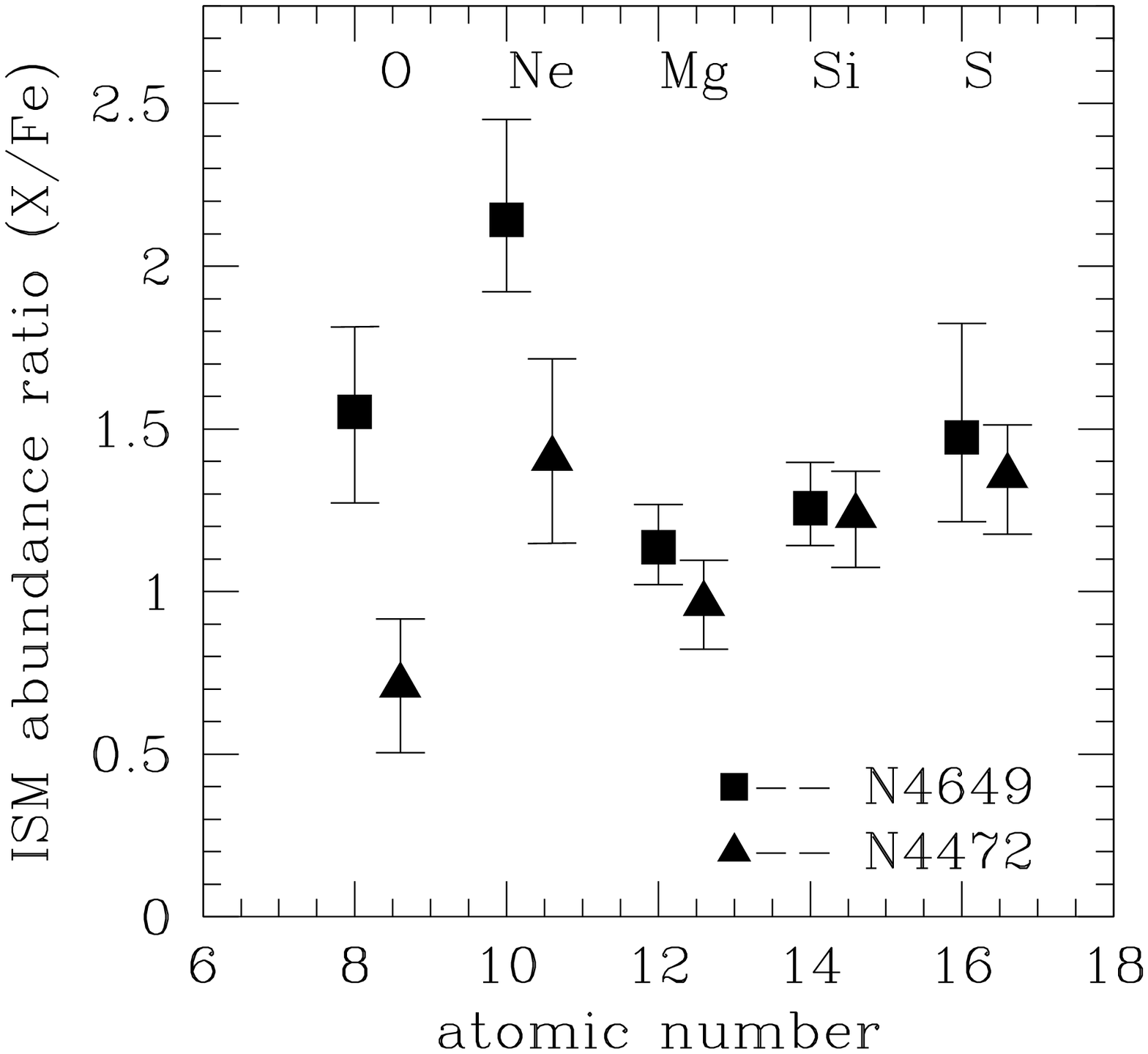}\hfil
\caption{{Abundance pattern comparison between NGC 4649
    (solid squares) and NGC 4472 (solid triangles). The NGC 4472
    values from Paper I have been rescaled to the \cite{agss09} solar
    standard.}}
\end{figure}

\subsection{\it XMM-Newton Spectral Analysis and Comparison with {\it
Suzaku}}

Compared to {\it Suzaku}, we found the {\it XMM-Newton} spectral fits
considerably more sensitive to the treatment of the (substantially
more prominent) background -- likely due to the presence of North
Polar Spur emission \citep{rsi04}. In the end, we adopt what may be
considered the most conservative approach -- confining the fit to the
0.3-5.5 keV band (MOS) or 0.4-5.5 keV band (PN) and, as detailed in
the previous section, following the {\it XMM-ESAS} prescription that
exclusively subtracts an estimate of the quiescent particle
background. The {\it XMM-ESAS} generation of instrumental backgrounds
does not remove the fluorescent spectral line features; and, some
residual soft proton contamination may also be present in the
extracted spectra. Thus we include the following additional NXB
components in our EPIC spectral models: narrow, fixed-energy Gaussian
emission lines corresponding to Al K$\alpha$ at 1.49 keV (MOS and PN)
and Si K$\alpha$ at 1.75 keV (MOS only) to account for the former;
and, a powerlaw with index fixed at 0.3 for the latter. As with the
XIS spectral fits we include thermal components, modeled with the {\bf
  apec} plasma code, to represent LHB, MWH, and ICM emission, and a
power law to account for the unresolved CXB. We again find that the
addition of a narrow 1.24 keV line improved fits in the inner region,
and this is included in $0-2'$ and $2-4'$ EPIC, as well as RGS, fits.

We display the best-fit model parameters for simultaneous EPIC MOS1,
MOS2, and PN $0-2'$ region spectral fits, and also for fits to the RGS
spectra, for the combined observations (2001 and 2007) in Table 5.
Spectra, best-fit models, and residuals are shown in Figure 8 for the
former, and in Figure 9 for the latter. We compare $0-2'$ {\it
  Suzaku}, $0-2'$ EPIC, and RGS abundance determinations in Figure 10,
finding good agreement in general and excellent agreement with respect
to the temperatures and Fe abundances.  A discrepancy in the intrinsic
absorption persists, with a column density consistent with the
Galactic value of 2.2$\times$10$^{20}$ cm$^{-2}$ \citep{dl90} for the
EPIC fits. We note that, other than the reduction in goodness-of-fit,
fixing the column density in fits to {\it Suzaku} spectra at the lower
Galactic value has little effect on the best-fit parameters -- the
largest effect is a $\sim 25$\% reduction in the $0-2'$ O
abundance. Fixing the column density in the $0-2'$ EPIC spectra at the
best-fit {\it Suzaku} value of $8.4\times 10^{20}$ cm$^{-2}$ does not
impact the goodness-of-fit and, likewise, shifts best-fit abundance by
$\sim 20$\% or less.  An even smaller effect is found if we restrict
the XIS fits to energies $>0.7$ keV and fix the column density at the
Galactic value.

\begin{figure}[ht]
\centering
\includegraphics[scale=0.33,angle=-90]{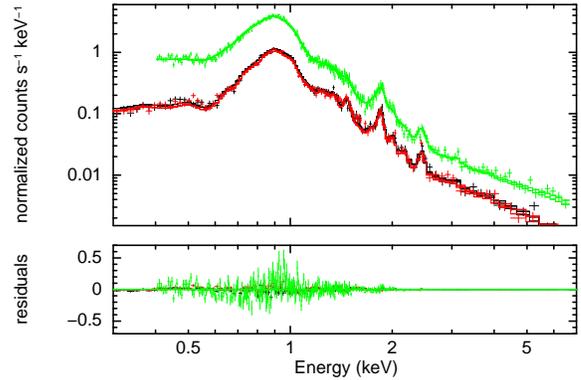}\hfil
\caption{{$0-2'$ 2001 {\it XMM-Newton} EPIC spectra
    (MOS1: black, MOS2: red, PN: green) and best-fit model with
    residuals (Table 5). The spectra are fit over the 0.3-5.5 keV band
    for the MOS and the 0.4-5.5 keV band for the PN. }}
\end{figure}

\begin{figure}[ht]
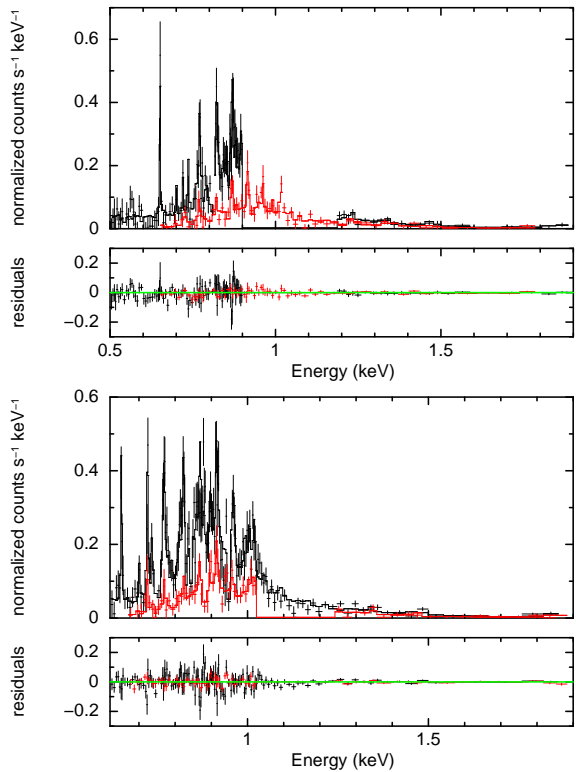

\centering
\includegraphics[scale=0.33,angle=-90]{n4649_r1_o12_050_nlinr.ps}\hfil
\includegraphics[scale=0.33,angle=-90]{n4649_r2_o12_050_nlinr.ps}
\caption{{RGS spectra (2007 Observation) and best-fit
    {\bf vapec} thermal plasma model (Table 5), and residuals. RGS-1
    ({\it top} panel, (a)) and RGS-2 ({\it bottom} panel, (b)) are
    broken out for clarity. First and second order spectra, included
    in the fitting, are shown.}}
\end{figure}

\begin{deluxetable*}{ccccccccccc}
\tabletypesize{\scriptsize} 
\tablewidth{0pt} \tablecaption{{\it XMM-Newton} RGS and EPIC $0-2'$
Best-Fit ISM Thermal Parameters}
\tablehead{\colhead{Spectra} & \colhead{$\chi^2/{\nu}$} &
    \colhead{$kT$} & \colhead{O} & \colhead{Ne} & \colhead{Mg} &
    \colhead{Si} & \colhead{S} & \colhead{Fe} & \colhead{Ni}}
\startdata 
EPIC (2001+2007) & 1756/1202 & $0.862_{-0.004}^{+0.011}$ & $0.64_{-0.06}^{+0.05}$ &
$1.52_{-0.12}^{+0.18}$ & $1.34_{-0.05}^{+0.04}$ &
$1.25_{-0.05}^{+0.04}$ & $1.41_{-0.12}^{+0.11}$ &
$0.97_{-0.01}^{+0.01}$ & $3.03_{-0.13}^{+0.14}$ \\
RGS (2001+2007) & 822/753 & $0.866_{-0.007}^{+0.020}$ &
$1.0_{-0.15}^{+0.15}$ & $1.66_{-0.33}^{+0.34}$ &
$0.94_{-0.27}^{+0.30}$ & $1.51_{-0.57}^{+0.60}$ & ... &
$0.97_{-0.08}^{+0.07}$ & $2.93_{-0.56}^{+0.62}$\\
\enddata 
\end{deluxetable*}

\begin{figure}[ht]
\centering
\includegraphics[scale=0.4,angle=0]{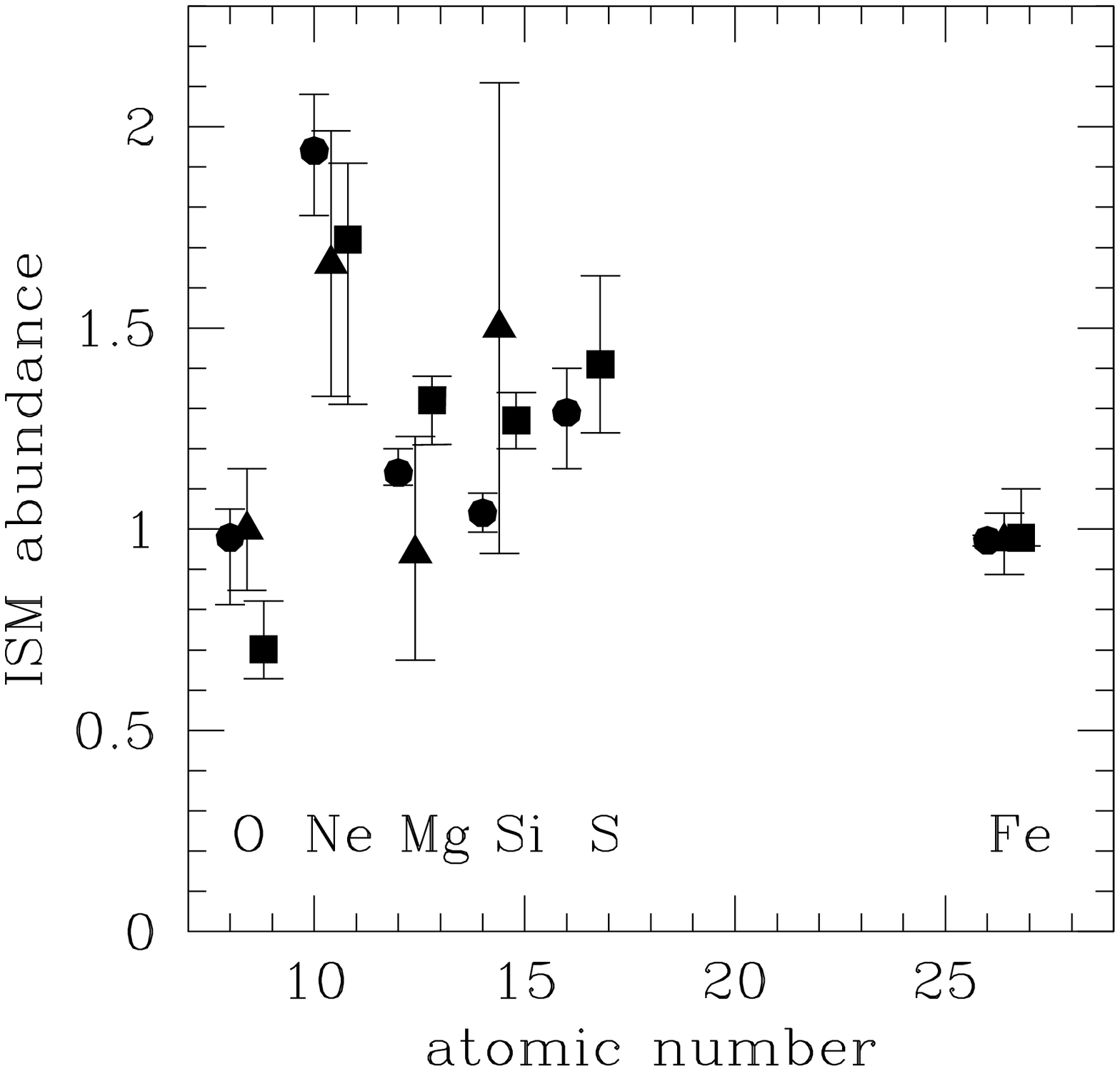}\hfil
\caption{{Abundance pattern comparison for separate fits
    to {\it XMM-Newton} EPIC $0-2'$ (solid squares) and RGS (solid
    triangles) and {\it Suzaku} XIS $0-2'$ (solid circles) spectra.}}
\end{figure}

The results of fitting the EPIC spectra extracted from the $0-2'$,
$2-4'$, and $4-6'$ annuli are displayed in Table 6. The higher
background in the EPIC cameras did not allow us to successfully fit
the 6-8$'$ annulus. Previous analysis of {\it XMM-Newton} data
\citep{rsi06,nm09,ji09} are generally in agreement with our results
although there are some significant discrepancies at some radii. As
noted above (Section 3; see, also, Appendix A), comparison is
complicated by the recent {\em AtomDB} update and, in fact, we find
fair agreement in the inner regions if we shift to the older {\em
  AtomDB} 1.3.2 version and include a second, hotter ISM component
(slightly favored in this case). Moreover, these studies employ
simpler and less conservative background treatments that utilize
subtraction of spectra extracted from ``off-source'' regions, while we
explicitly include GXB and ICM components, allowing for their local
variation and that of the instrumental background. For example, our
central Fe and O abundances in the inner $2'$ agree well with previous
work while we find lower Si abundance than \citep{nm09,ji09} in the
same region. In the $2-4'$ and $4-6'$ rings, where the background is
more significant, we find higher abundances than these previous works.

\begin{deluxetable*}{ccccccccc}
\tabletypesize{\scriptsize}
\tablewidth{0pt}
\tablecaption{{\it XMM-Newton} Annular Best-Fit ISM Parameters}
\tablehead{\colhead{region} & \colhead{$\chi^2/{\nu}$} &
  \colhead{$kT$} & \colhead{0} & \colhead{Ne} &
  \colhead{Mg} & \colhead{Si} & \colhead{S} & \colhead{Fe}}
\startdata

{$0-2'$} & 1756/1202 & $0.862_{-0.004}^{+0.011}$ & $0.64_{-0.06}^{+0.05}$ &
$1.52_{-0.12}^{+0.18}$ & $1.34_{-0.05}^{+0.04}$ &
$1.25_{-0.05}^{+0.04}$ & $1.41_{-0.12}^{+0.11}$ &
$0.97_{-0.01}^{+0.01}$ \\

{$2-4'$} & 1114/910 & 
$0.966_{-0.015}^{+0.008}$ & $1.14_{-0.11}^{+0.11}$ &
$1.23_{-0.24}^{+0.71}$ & $1.29_{-0.19}^{+0.08}$ &
$1.46_{-0.11}^{+0.08}$ & $1.27_{-0.27}^{+0.22}$ &
$1.09_{-0.03}^{+0.01}$\\

{$4-6'$} & 809/621 & $0.960_{-0.045}^{+0.032}$ &
$0.50_{-0.16}^{+0.16}$ & $0.86_{-0.41}^{+0.43}$ &
$0.31_{-0.20}^{+0.22}$ & $0.70_{-0.12}^{+0.21}$ &
$0.99_{-0.36}^{+0.41}$ & $0.45_{-0.09}^{+0.12}$\\

\enddata 

\tablecomments{Displayed for each spectral extraction region are
  minimized $\chi^2$/dof for simultaneous fits to {\it XMM-Newton}
  EPIC spectra. Shown are best-fit ISM temperatures (in keV), and
  best-fit O, Ne, Mg, Si, S, and Fe abundances referred to the solar
  standard of \cite{agss09}. The column density is fixed at the
  Galactic value, 2.2$\times 10^{20}$ cm$^{-2}$.}
\end{deluxetable*}

We plot the ratio of the EPIC-derived abundances to those of {\it
  Suzaku} in Figures 11abc, and the EPIC abundance ratio profiles
(with respect to Fe) in Figures 12ab. The latter show a dropoff in the
$4-6'$ annulus with respect to the inner $4'$. The EPIC S/Fe and Si/Fe
(Figure 11b) profiles echo those seen in the XIS data (Figure 6b) in
indicating positive gradients. However, while {\it XMM-Newton} and
{\it Suzaku} abundances show good agreement in the $0-2'$ and $4-6'$
regions, the EPIC abundances are more discrepant in the $2-4'$
annulus. We have investigated this annulus in more detail by
conducting joint EPIC/XIS fits, considering different bandpasses, and
examining fits with the absorption column density both free and fixed.
In doing so we found that the EPIC spectra admit of two fits of
comparable quality -- one with Galactic absorption and relatively high
abundances (Table 6), one with elevated absorption and lower
abundances that is in accord with the {\it Suzkau} fits (Table
4). However the converse is not true: the XIS spectra are not
consistent with the higher abundance model.

\begin{figure}[ht]
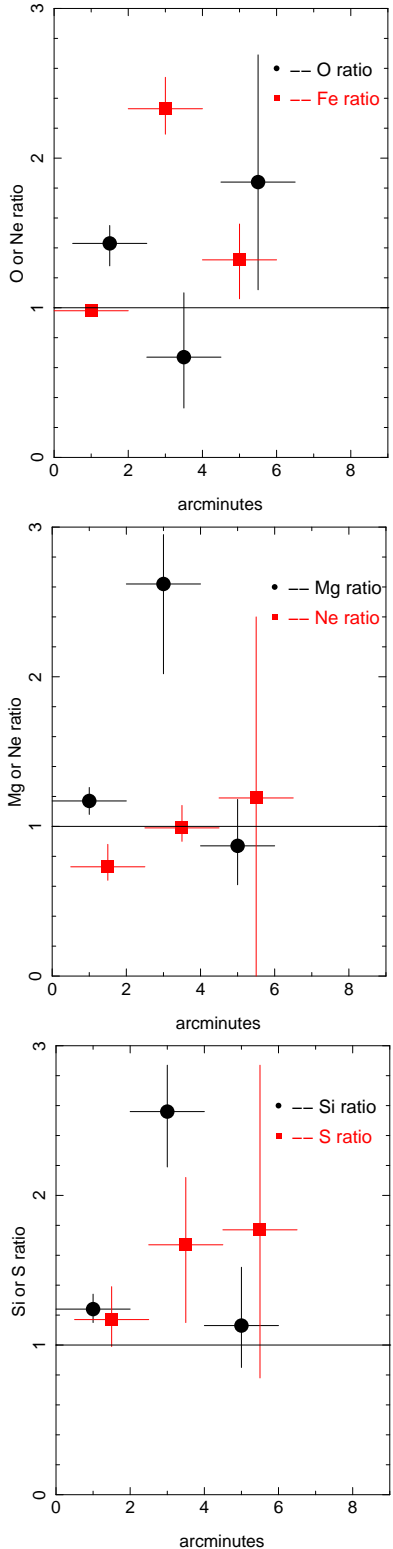

\centering
\includegraphics[scale=0.28,angle=0]{xmm_Suzaku_OFeratio_radius.ps}\hfil
\includegraphics[scale=0.28,angle=0]{xmm_Suzaku_MgNeratio_radius.ps}\hfil
\includegraphics[scale=0.28,angle=0]{xmm_Suzaku_SiSratio_radius.ps}
\caption{{Ratio of the abundances derived from the {\it
      XMM-Newton} EPIC data to those of the {\it Suzaku} XIS
    detectors. We show Fe and O ({\it top} panel, (a)), Si and S ({\it
      middle} panel, (b)), Mg and Ne ({\it bottom} panel, (c)) ratio
    profiles. The points represented by circles are shifted to the
    right by $0.5'$.}}
\end{figure}

\begin{figure}[ht]
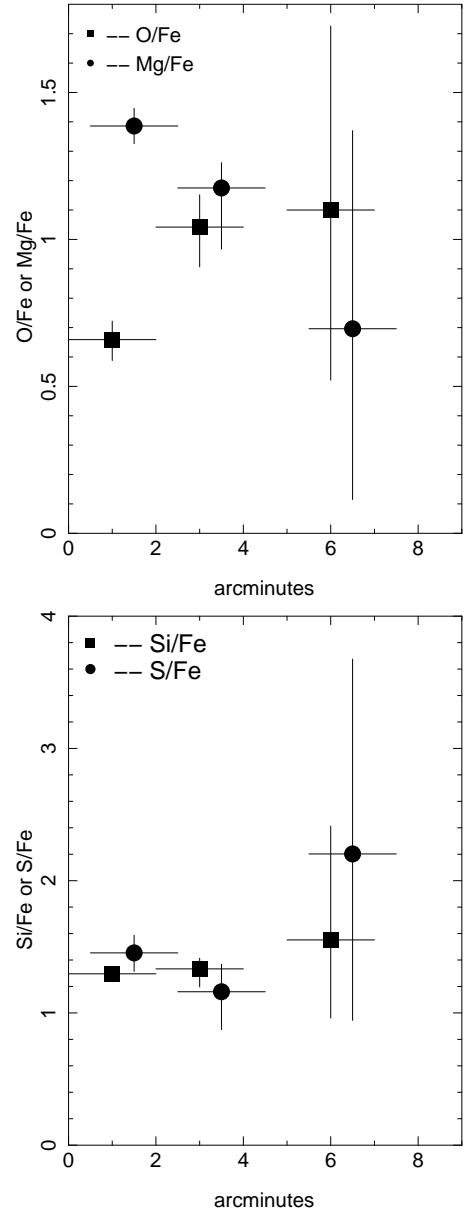

\centering
\includegraphics[scale=0.33,angle=0]{O_Mg_Ferat_epic.ps}\hfil
\includegraphics[scale=0.33,angle=0]{Si_S_Ferat_epic.ps}
\caption{{Profiles of abundance ratios with respect to
    Fe for O and Mg ({\it top} panel, (a)); and, Si and S ({\it
      bottom} panel, (b)) derived from the EPIC data.}}
\end{figure}

Based on the above considerations, we adopt the abundances that we
infer from fitting the {\it Suzaku} spectra for comparison with simple
chemical evolution models. These are evidently more robust, for the
crude spatial binning that we employ, because of the lower background
and sharper energy resolution of {\it Suzaku} (see Paper I). As a
result we extend the measured abundance pattern to greater radii,
encompassing a larger fraction of the optical galaxy than was
previously possible. A parallel analysis that utilizes the {\it
  XMM-Newton} results in the same qualititative results and
conclusions.

\section{Modeling the NGC 4649 Abundance Pattern}

Although the host of a weak AGN \citep{shu08,ho09,dunn10}, NGC 4649 is
not strongly disturbed optically \citep{fer06} or in X-rays
\citep{hum08} and has an old, passively evolving stellar population
\citep{jtb06}. However, there are strong indications for complex
time-dependent circulation flows in the NGC 4649 ISM. On one hand, the
thermodynamic characteristics of the gas within 150 pc and the
flattened isophotes further out inferred from {\it Chandra} provide
evidence for inflow from kpc scales all the way into the $3\times
10^9$ M$_{\sun}$ \citep{hum08} nuclear black hole
\citep{ds07,bmhb09}. On the other, the fact that this material is
rapidly cooling, yet has not appreciably accumulated in the core (in
gaseous or stellar form), implies the existence of an outward
transport mechanism that is probably intermittent \citep{bmhb09}. The
impact of these flows on the integrated ISM abundances depends on the
metallicity in the flow and its ultimate disposition -- in particular
whether it escapes the galaxy entirely and permanently.

We utilize the abundance pattern derived from fits to the {\it Suzaku}
spectra in constraining the enrichment of the hot ISM in NGC 4649. We
adopt a global perspective in our interpretation of the NGC 4649
abundance pattern by seeking explanation of the hot ISM abundances
measured over the entire optical galaxy via X-ray spectroscopy as
derived in previous sections, and defer detailed consideration of
abundance, and possible abundance ratio, gradients. This is partially
driven by the limitations of the {\it Suzaku} angular resolution, and
the attendant degeneracies and uncertainties in spectral modeling, as
well as the relative robustness of the galaxy-wide emission averaged
abundances (and ratios). Our goal is to examine the overall sources
and sinks of the metals measured in this way, and attempt to quantify
their relative contributions. In this way, we may identify signatures
of the metal transport and enrichment mechanisms discussed above, and
better understand the ``ins and outs'' \citep{mat90} of the gas flow
in this elliptical galaxy.

\subsection{Conservation Equations and Steady-State Solution}

We proceed with a chemical evolution formalism along the lines of that
in Paper I, with the notable additions of explicitly allowing for
inflow, and the inclusion of recently published SNIa yields
\citep{mae10}. The equations for the conservation of overall ISM mass,
$M_{\rm ISM}$, and mass of the $ith$ element, $f^i_{\rm ISM}M_{\rm
ISM}$, in a fully formed elliptical galaxy are as follows:

\begin{equation}
{{dM_{\rm ISM}}\over {dt}}=\dot M_{\rm MR}-\dot M_{\rm out}+\dot
M_{\rm in},
\end{equation}

and

\begin{equation}
{{d(f^i_{\rm ISM}M_{\rm ISM}})\over {dt}}=G^i_{\rm SNIa}+G^i_{\rm
MR}-(1+b^i_{\rm out})\dot M_{\rm out}f^i_{\rm ISM}+\dot M_{\rm
in}f^i_{\rm in},
\end{equation}
where $f^i_{\rm ISM}$ is the mass fraction, $\dot M_{\rm MR}$ and
$G^i_{\rm MR}$ are the mass injection rates for the total gas mass and
for the $ith$ element from evolved stars, and $G^i_{\rm SNIa}$ the
metal mass injection rate from SNIa for the $ith$ element. $G^i_{\rm
  MR}=\dot M_{\rm MR}f^i_{\rm stars}$, where $f^i_{\rm stars}$ is the
mass fraction of the $ith$ element in these mass-losing stars, and
$G^i_{\rm SNIa}=\dot N_{\rm SNIa}y_{\rm SNIa}^i$, where $\dot N_{\rm
  SNIa}$ is the SNIa rate and $y_{\rm SNIa}^i$ is the $ith$ element
SNIa mass yield. $\dot M_{\rm out}$ ($\dot M_{\rm in}$) is the rate of
flow out of (into) the system, $b^i_{\rm out}$ a bias factor allowing
for an abundance offset for the outflowing gas (Paper I), and
$f^i_{\rm in}$ the mass fraction of the $ith$ element in the inflowing
gas. $\dot M_{\rm out}$ refers to the flow of gas beyond $\sim
6.5R_e$, $\dot M_{\rm in}$ to the introduction of gas into this same
volume by any means other than that associated with stellar mass loss.
For a single-phase outflow one expects $b^i_{\rm out}<0$ as a
result of negative abundance gradients in the flow, and $b^i_{\rm
out}>0$ if gas of higher metallicity preferentially exits the galaxy
due to buoyancy effects in hot SNIa ejecta \citep{tw10}. For
simplicity we assume that $b^i_{\rm out}=b_{\rm out}$ is identical for
all elements (Paper I), and that the abundances in the inflow are in
the same proportions as in the stars, $f^i_{\rm in}=\phi_{\rm
in}f^i_{\rm stars}$.

In the present work we consider equations (1) and (2) in the steady
state limit, in which case the source terms may be considered as
averages over the density-weighted-mean residence time of the
gas. Solving for the ISM abundances,
\begin{equation}
f^i_{\rm ISM}={{f^i_{\rm stars}(1+r_{\rm in}\phi_{\rm in})+qy_{\rm
SNIa}^i}\over {(1+b^i_{\rm out})(1+r_{\rm in})}},
\end{equation}
where $r_{\rm in}\equiv \dot M_{\rm in }/\dot M_{\rm MR}=\dot M_{\rm
  out}/\dot M_{\rm MR}-1$ and $q=0.00667\theta_{\rm SNIa}/\theta_{\rm
  MR}~M_{\sun}^{-1}$. $\theta_{\rm MR}$ is the specific mass loss rate
in units of $2.4\times 10^{-11}~M_{\sun} {\rm yr}^{-1}~L_{B\sun}^{-1}$
(Paper I), and $\theta_{\rm SNIa}$ is the specific SNIa rate in units
of the estimated present-day elliptical galaxy rate of 0.16
SNU\footnote{SN per century per $10^{10}~L_{B\sun}$}
\citep{cap97,crt99,dil08}.

\subsection{Source Terms}

ISM abundances depend on the ratio $\theta_{\rm SNIa}/\theta_{\rm MR}$
through the parameter $q$ in equation (3). We generally assume
$\theta_{\rm MR}=1$ when quoting values of $\theta_{\rm SNIa}$ based
on observationally inferred values of $q$. The corresponding stellar
mass return rate (at the present time), derived in Paper I, is $\sim
40-60$\% times higher than often assumed based on older estimates of
turnoff and stellar remnant masses \citep{fg76,mat89,cpr91}. $\dot
M_{\rm MR}$ decreases with time approximately as $t^{-1.3}$ (Paper I),
so that the effective value of $\theta_{\rm MR}$ may be greater if
some portion of the ISM was lost from stars at an earlier epoch -- the
average over the past 10 Gyr is 2.2 times the present value. At the
same time, estimates of the number of SNIa produced per stellar mass
formed ($\sim 0.0005~M_{\sun}^{-1}$; e.g., Horiuchi and Beacom 2010)
result in an average SNIa rate three times our fiducial rate of 0.16
SNU (though $\sim$ half of these may be prompt explosions that are
irrelevant for enriching the present-day ISM). Also, a recent estimate
of the SNIa rate \citep{li11} is $\sim$twice our fiducial value, and
there is evidence of an enhanced rate in clusters
\citep{man08,dil10,msg10}. Further complication arises due to the
absence of a compelling theoretical explanation for rates as high as
these \citep{rbf09,men10}. In other words, $\theta_{\rm
SNIa}=\theta_{\rm MR}=1$ represents a reasonable benchmark but not a
strict expectation: the larger values estimated from some optical
surveys lack a firm theoretical basis that would indicate the
magnitude and evolution of the SNIa rate in any individual elliptical
galaxy with its particular structure and stellar population.

In Paper I, individual stellar abundances are determined with
reference to a specific SNII yield set \citep{kob06} and the
assumption that the SNIa enrichment of the stellar and ISM components
are identical; the SNIa/SNII mix was then related to the stellar
$\alpha/Fe$ ratio that was allowed to vary. Here we adopt the simpler
assumption that stellar O, Ne, Mg, Al, Si, S, and Ar abundances are in
solar proportions and overabundant with respect to Ca, Fe, and Ni by a
fixed, constant factor given by the $\alpha/Fe$ ratio. We base this on
the consistency we found in Paper I between the optically measured
$[\alpha/Fe]_{\rm stars}$ and those in our best-fit models to the hot
ISM abundance pattern in NGC 4472. Moreover, there is no consensus as
to which of the many available SNII yield sets (if any) is appropriate
\citep{glm97,rom10}, especially given the uncertainties in, e.g., the
initial mass function of the original stellar population and the
possibility that SNII yields vary as the metallicity of the stellar
population that provides the Type II progenitors accumulates. There is
also the possibility that the primordial enrichment of the stars and
the ongoing enrichment of the hot ISM are characterized by distinct
yields from separate components of a bimodal populations of SNIa
\citep{mdp06,mao11,hb10} or, as may be the case with SNII, that SNIa
yields evolve with the progenitor population -- in their case due to
effects of age and metallicity on the deflagration-to-detonation
transition density \citep{bra10,kru10,jac10}.

In this way, we focus on the period subsequent to the epoch of
formation of the bulk of the stars, and only assume that an old
stellar population with the optically estimated $[\alpha/Fe]_{\rm
stars}$ has emerged during the unspecified prior evolution. However,
one must keep in mind that there is evidence that individual
$\alpha$-elements deviate from solar ratios in massive ellipticals,
and do not vary in lockstep in general (Graves et al. 2007, Smith et
al. 2009, and references therein) -- in fact our grouping of Ca with
Fe above is based on just such evidence for that element (see, also,
Pipino \& Matteucci 2004; de Plaa et al. 2007). Following
\cite{hb06,hum08}, we estimate the following global stellar values
from measurements of optical line indices
\citep{tra00a,tra00b,tho05,how05,san06}: $[Fe/H]_{\rm stars}=0\pm 0.2$
and $[\alpha/Fe]_{\rm stars}=0.25\pm 0.05$, and adopt
$[\alpha/Fe]_{\rm stars}=0.25$. The only relevant supernova yields are
those of SNIa that explode after long ($>$Gyr) evolutionary delay
times. We consider a wide range of published SNIa yield sets,
including those derived from the W7 \citep{nom97} and more recent
C-DEF \citep{mae10} pure deflagration models, and those based on
one-dimensional delayed detonation (DD) models with a range of central
and transition densities \citep{nom97,iwa99}. For the first time, we
consider the yields based on the two- dimensional DD models of
\cite{mae10}, including their off-center DD model (O-DDT).

Finally, in comparing our models with the NGC 4649 ISM abundance
pattern we consider only elements that have distinct and well-measured
emission features in the X-ray spectra -- O, Mg, Si, S, and Fe --
although we also plot Ne when displaying this comparison.

\subsection{Solutions Applied to NGC 4649}

We consider two classes of simplified solutions (with neither outflow
nor inflow) to equation (3) for illustrative purposes. In the first
the SNIa rate is set to 0 ($q=0$) so that ISM abundances are
proportional to stellar abundances. Figure 13a includes a comparison
of the hot ISM and stellar abundance patterns for solar and half-solar
stellar Fe abundances ($[Fe/H]_{\rm stars}=0$, -0.3). Also displayed
is the best-fit model including SNIa with $\theta_{\rm
  SNIa}/\theta_{\rm MR}=1$: the $[Fe/H]_{\rm stars}=-0.3$ C-DEF model
(where SNIa yields for relevant elements are lowest). Notably,
abundances are generally overpredicted -- even for the $[Fe/H]_{\rm
  stars}=-0.3$ model with no SNIa.

We now turn to more general steady state solutions. Once the stellar
abundance ratios and SNIa yields are specified, the ISM abundance
ratios depend on a single parameter. Adopting Fe as a reference
element, the ratios are determined by the following:
\begin{equation}
{{f^i_{\rm ISM}}\over {f^{\rm Fe}_{\rm ISM}}}= {{{f^i_{\rm
stars}}/{f^{\rm Fe}_{\rm stars}} +py_{\rm SNIa}^i}\over {1+py_{\rm
SNIa}^{\rm Fe}}},
\end{equation}
for each element $i$, where ${{f^i_{\rm stars}}/ {f^{\rm Fe}_{\rm
stars}}}$ is the corresponding stellar ratio (the solar ratio for Ca
and Ni, and 1.8 times the solar ratio otherwise; see above), and
$p\equiv q(1+r_{\rm in}\phi_{\rm in})^{-1}/{f^{\rm Fe}_{\rm
stars}}$. Essentially, this expresses the simple dependence of the ISM
ratios on the ratio of SNIa-to-stellar direct injection, modified by
inflow (for $r_{\rm in}>0$); a uniformly biased outflow does not
affect the ratios.

We determine the best-fit (minimum $\chi^2$) value of $p$ to the
observed NGC 4649 abundance ratios for each yield set. Figure 13b
shows the corresponding abundance patterns for the more recently
published yields compared to the observations. Many of the yield sets
provide good fits to the abundance ratios (as does the W7 model); the
exception being the \cite{mae10} C-DDT model with its exceptionally
high Si-to-Fe ratio. We do not find evidence for the oft-reported
underabundance of O (see Paper I, and references therein) -- in part
because of the lower solar standard O abundances in \cite{agss09} that
we adopt.

\begin{figure}[ht]
\centering
\includegraphics[scale=0.4,angle=0]{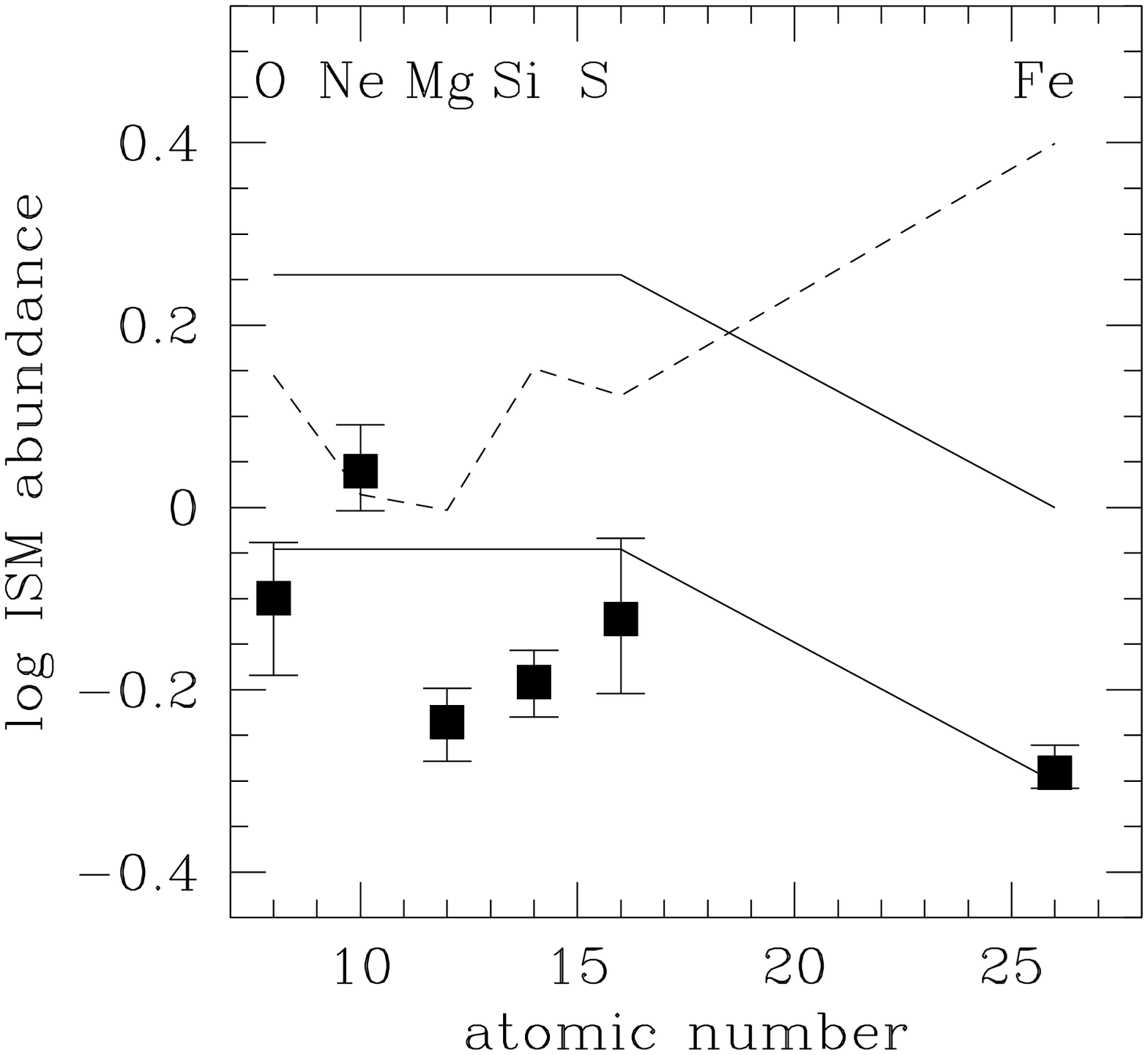}\hfil
\includegraphics[scale=0.4,angle=0]{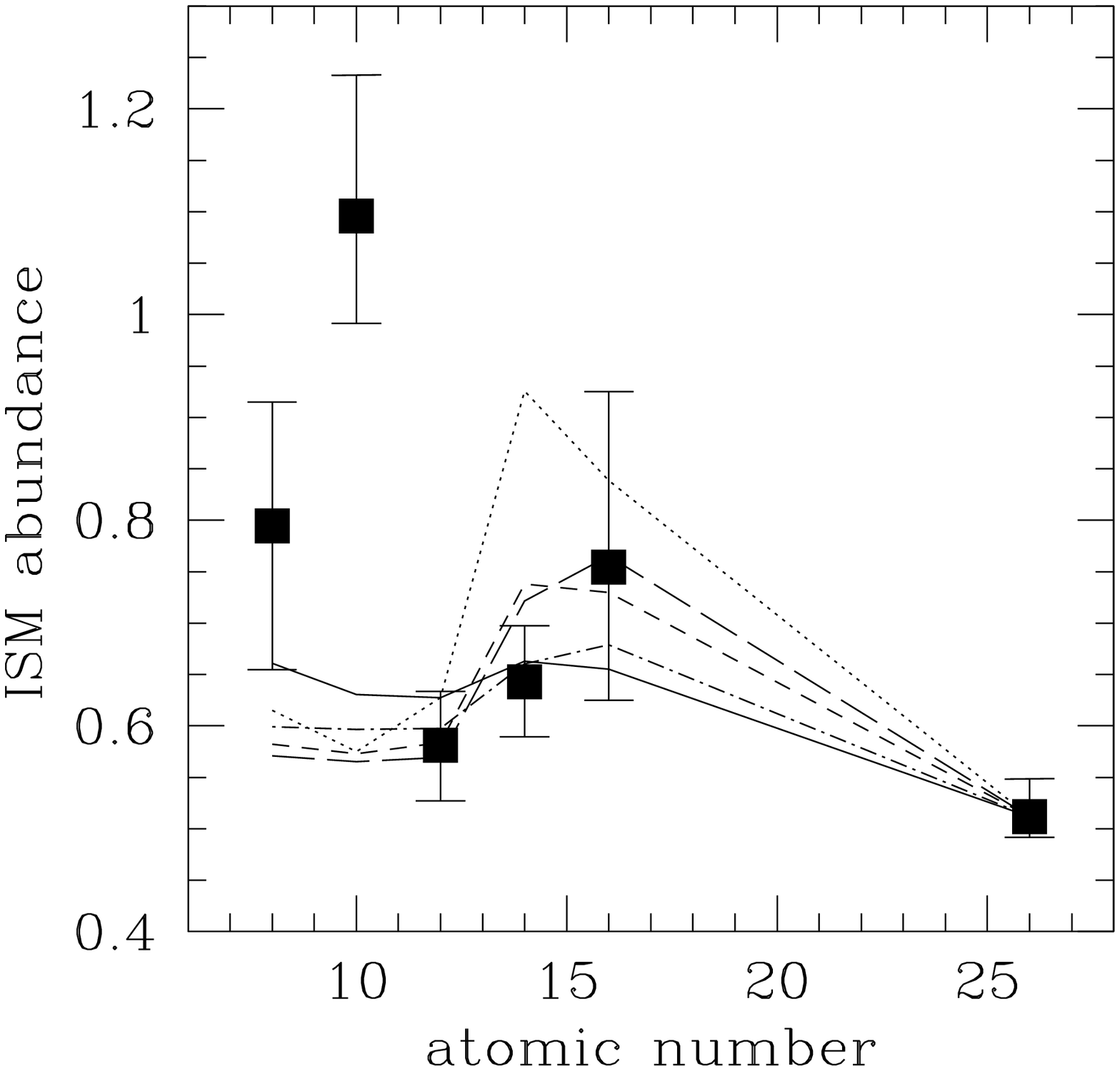}
\caption{{{\it Top} panel (a): Comparison of NGC 4649
    abundance pattern (filled squares) and three simple (no outflow or
    inflow) model predictions: stars-only enrichment with $[Fe/H]_{\rm
      stars}=0$ and $-0.3$ (upper and lower solid lines), and the
    best-fit model with $[Fe/H]_{\rm stars}=-0.3$ and SNIa exploding
    at the standard rate (C-DEF yields, broken line). {\it Bottom}
    panel (b): Best-fit solutions to equation (4) to the NGC 4649
    abundance ratio pattern (excluding Ne) for the following SNIa
    abundance sets: \cite{mae10} C-DEF ($p=1.25$), C-DDT ($p=1.72$),
    and O-DDT ($p=0.94$) models (solid, dotted, and short-dashed
    lines), \cite{iwa99} WDD1 ($p=0.94$) and WDD3 ($p=0.63$) models
    (long-dashed, dot-dashed lines; the WDD2 solution closely
    resembles that for the O-DDT yields). The solutions, which predict
    only relative abundances, are scaled to the observed ISM Fe
    abundance.}}
\end{figure}

\subsection{Discussion}

\subsubsection{Astrophysical Interpretation of Successful Models}

How do the dimensionless model solutions that match the observed hot
ISM abundances translate into astrophysical quantities of interest,
and are these reasonable? We address that in this subsection by
unfolding the single model parameter $p$ into the implied range in the
combination of inflow, outflow, and SNIa properties in NGC 4649. The
value of $p$, determined by the observed abundance ratios, determines
the relation between the inflow term $r_{\rm in}\phi_{\rm in}$ and the
SNIa-to-stellar-mass-return ratio term $q$ (see above), and sets a
minimum value of $q$ ($q_{\rm min}$) such that $r_{\rm in}\phi_{\rm
  in}>0$: $q_{\rm min}=pf^{\rm Fe}_{\rm stars}$. If one then specifies
$f^{\rm Fe}_{\rm ISM}$ one may separately derive $r_{\rm in}$ and
$\phi_{\rm in}$ for a given outflow bias parameter, $b_{\rm out}$. The
minimum value of $q$ required for $r_{\rm in}>0$ {\it and} $\phi_{\rm
  in}>0$ is $q_{\rm min}=max[pf^{\rm Fe}_{\rm stars},(1+b_{\rm
    out})pf^{\rm Fe}_{\rm ISM}/(1+py_{\rm SNIa})]$. In what follows we
set $\theta_{\rm MR}=1$ when quoting values of $\theta_{\rm SNIa}$
(i.e., $\theta_{\rm SNIa}=150q$), and assume $[Fe/H]_{\rm stars}=0.0$,
unless otherwise specified. For illustrative purposes we adopt a
composite model with equal contributions from C-DEF, O-DDT, and WDD3
SNIa (the three distinct well-fitting models in Figure 13b); the
following discussion is qualitatively unaltered, and the
goodness-of-fit similar, for any of the individual well-fitting
models. The abundance pattern inferred from optimizing the match to
the data over the parameter $p$ for this composite model is shown in
Figure 14, where it is compared with the pure stellar and SNIa ratios,
as well as the predicted ratio for $\theta_{\rm SNIa}=1$ and no inflow
($r_{\rm in}\phi_{\rm in}=0$). The theoretical pattern is extended to
the additional elements C, N, Al, Cr, Mn, and Ni (stellar abundances
of the former three are assumed equal to those of O-through-Ar, the
latter three those of Ca and Fe; see Section 4.1.2). Figure 15a plots
$r_{\rm in}\phi_{\rm in}$ as a function of $\theta_{\rm SNIa}$ for
this optimum composite model (solid line in Figure 14). The minimum
SNIa rate ($r_{\rm in}\phi_{\rm in}\rightarrow0$), set by the lower
value of $[\alpha/Fe]$ in the ISM compared to the stars, corresponds
to $\theta_{\rm SNIa,min}=0.18$ (0.091 for $[Fe/H]_{\rm stars}=-0.3$);
more generically $\theta_{\rm SNIa}>0.12$ for all of the well-fitting
individual SNIa yield sets (for $[Fe/H]_{\rm stars}=0.0$). The curves
in Figure 15a may be expressed as $\theta_{\rm SNIa}=\theta_{\rm
  SNIa,min}(r_{\rm in}\phi_{\rm in}+1)$. That is, very large inflow
rates are required to reconcile the NGC 4649 ISM abundances with
$\theta_{\rm SNIa}=1$.

In Figure 15b, $r_{\rm in}$ and $\phi_{\rm in}$ are separately broken
out based on the measured $\sim$half-solar global value of $f^{\rm
  Fe}_{\rm ISM}$ derived from the {\it Suzaku} spectral fits. The
inferred abundances in the accreted gas are $<0.28$ times the (nearly
solar) stellar value for $\theta_{\rm SNIa}<1$, reasonable for an
intracluster or intragroup medium -- and consistent with the outermost
{\it Suzaku} annulus ($6-8'$) abundance (Table 3), while $r_{\rm
  in}>2.1$. The introduction of a biased outflow that preferentially
carries metals out of the galaxy naturally reduces the require inflow
rates. For $b_{\rm out}=1$ (abundances in the outflow twice that in
the stars), the inflow metallicity simply shifts to lower inflow rates
(Figure 15b) -- for fixed $\theta_{\rm SNIa}<1$ $r_{\rm in}\rightarrow
(r_{\rm in}-1)/2$; more generally, $r_{\rm in}\rightarrow (r_{\rm
  in}-b_{\rm out})/(1+b_{\rm out})$ (with a compensating increase in
$\phi_{\rm in}$ since $r_{\rm in}\phi_{\rm in}$ is independent of
$b_{\rm out}$). For $[Fe/H]_{\rm stars}=-0.3$ the required level of
inflow enrichment is higher for a given SNIa rate (Figure 15a) --
$r_{\rm in}$ is independent of $[Fe/H]_{\rm stars}$; for $[Fe/H]_{\rm
  stars}\rightarrow [Fe/H]_{\rm stars}'$ $\phi\rightarrow [(r_{\rm
    in}\phi_{\rm in}+1)([Fe/H]_{\rm stars}'/[Fe/H]_{\rm
    stars})-1]r_{\rm in}^{-1}$.

The relative importance of dilution is expected to increase with
radius as the timescale to replenish the ISM via stellar mass
increases, with part of this dilution perhaps arising from an {\it ab
  initio} group-like orgin of the gas there.  We do not account for
this, but will address this with future multizone models.

\begin{figure}[ht]
\centering
\includegraphics[scale=0.4,angle=0]{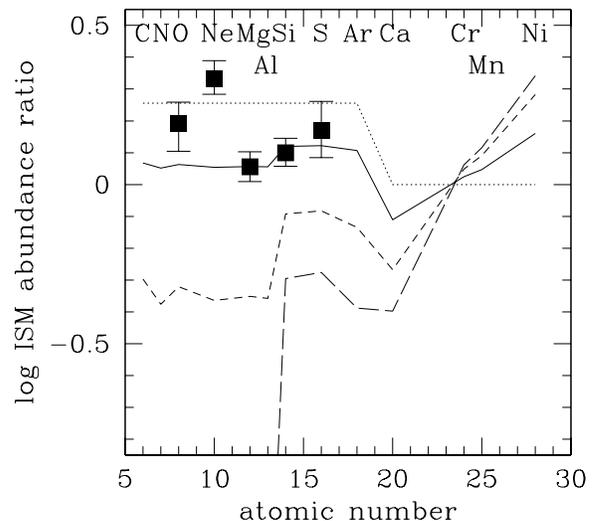}\hfil
\caption{{Best-fit model to the observed abundance
    pattern derived from {\it Suzaku} data (filled squares) assuming
    equal numbers of C-DEF, O-DDT, and WDD3 SNIa (solid line),
    compared with the ratios for pure stellar (dotted) and SNIa
    (long-dashed) enrichment and that for $\theta_{\rm SNIa}=1$,
    $r_{\rm in}\phi_{\rm in}=0$ (standard SNIa rate, no inflow;
    short-dashed line).}}
\end{figure}

\begin{figure}[ht]
\centering
\includegraphics[scale=0.4,angle=0]{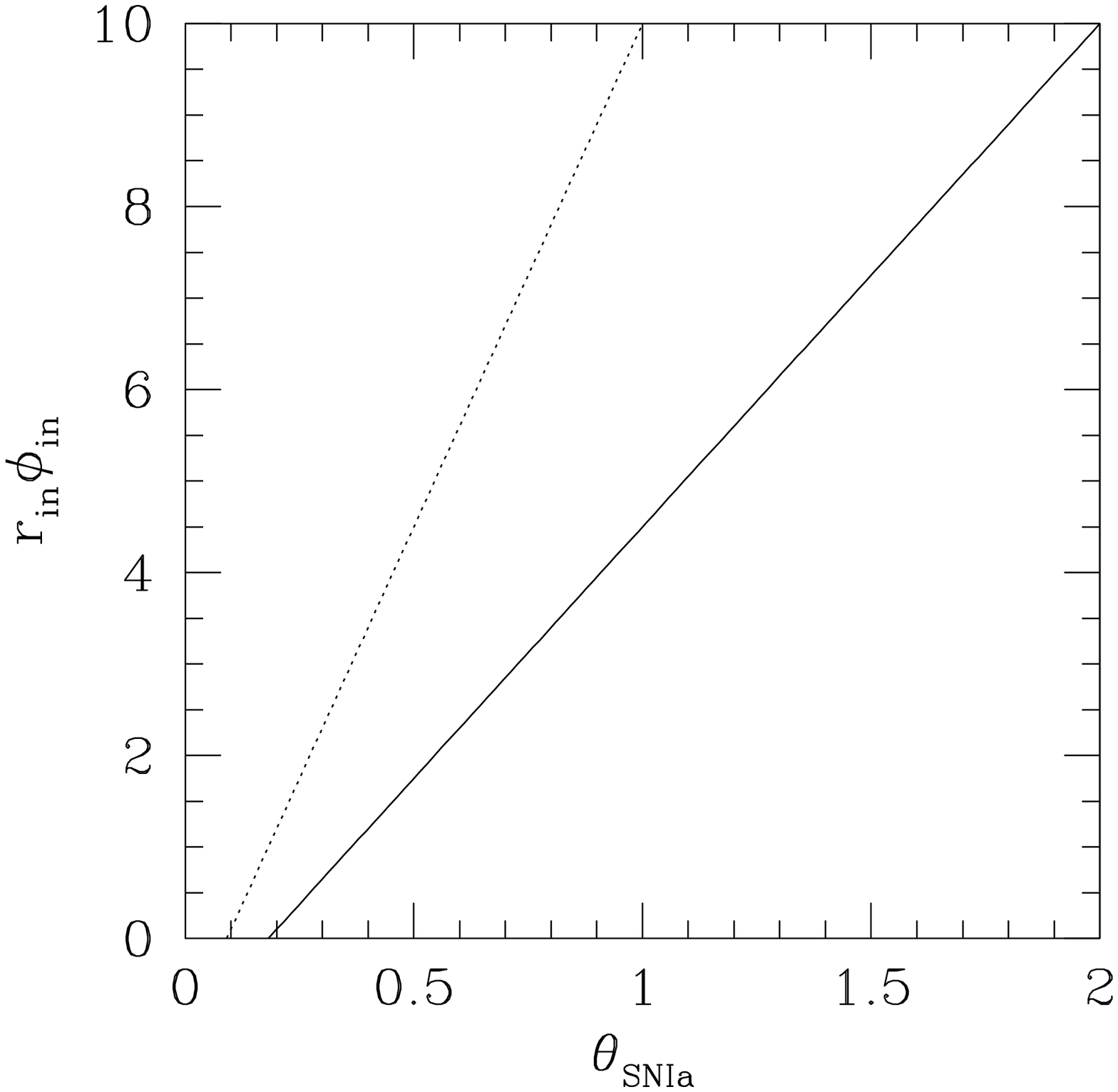}
\includegraphics[scale=0.4,angle=0]{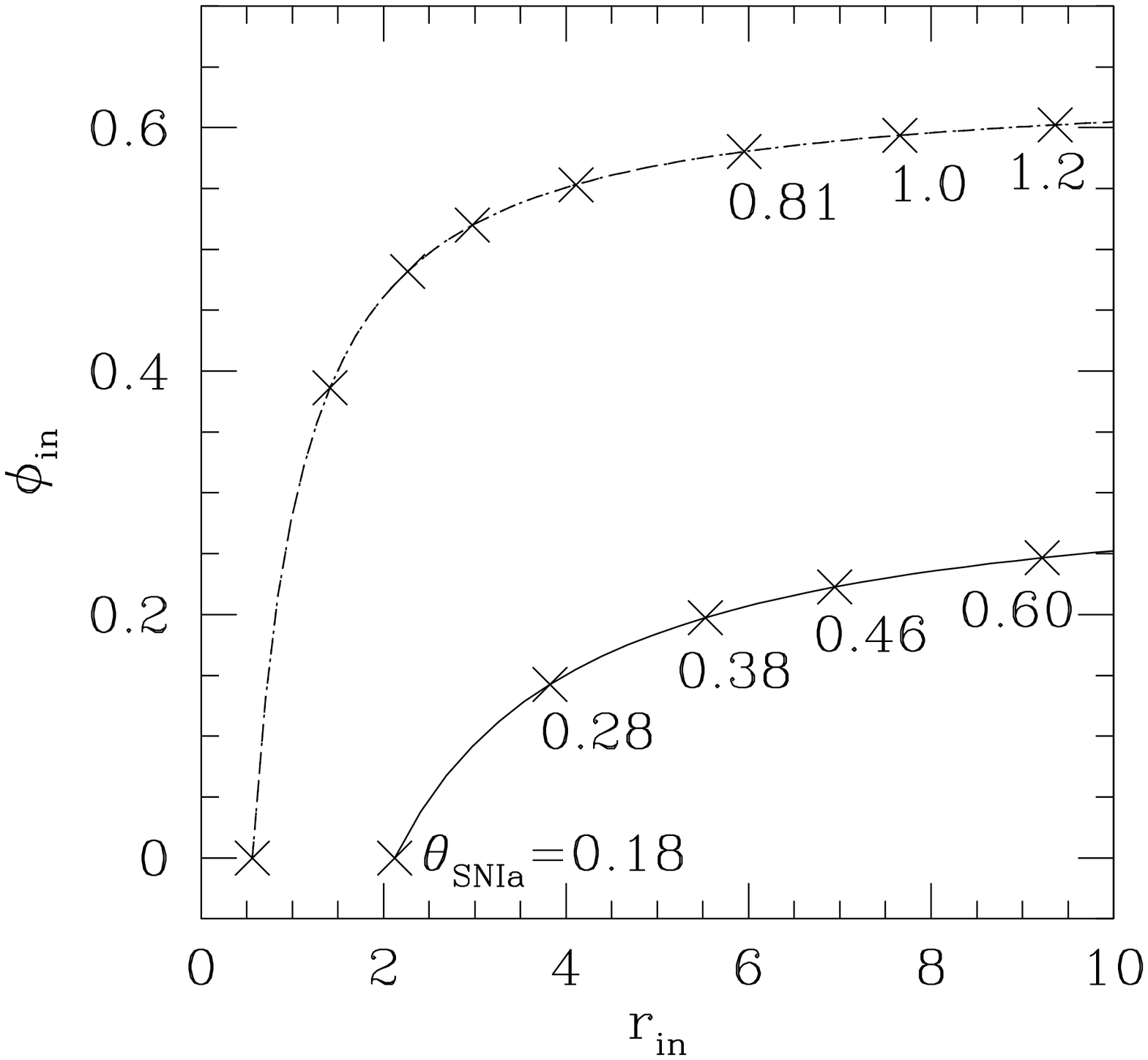}\hfil
\caption{{{\it Top} panel (a): $r_{\rm in}\phi_{\rm in}$
    vs. $\theta_{\rm SNIa}$ for the best-fit model (solid line in
    Figure 14). Also shown, is the corresponding curve for
    $[Fe/H]_{\rm stars}=-0.3$ (dotted line).} {\it Bottom} panel (b):
  $\phi_{\rm in}$ vs. $r_{\rm in}$ implied by the observed $f^{\rm
    Fe}_{\rm ISM}$ for the best-fit abundance ratio patterns assuming
  $[Fe/H]_{\rm stars}=0$ and $b_{\rm out}=0.0$ (solid curve) or 1.0
  (broken curve). The corresponding $\theta_{\rm SNIa}$ are shown by
  crosses; the first five values are the same for each curve. {\it
    Abundance ratios} are invariant along the lines in (a); {\it
    abundances} are invariant along the curves in (b) for the
  specified SNIa rates.}
\end{figure}

\subsubsection{NGC 4472 and Other Ellipticals (Briefly) Revisited}

As noted above (Figure 10), the abundance {\it ratio} patterns are
generally consistent between NGC 4472 and NGC 4649. This suggests that
the higher abundance of Fe in the former is not simply due to a higher
SNIa rate -- which would skew the abundance pattern -- but, in the
context of the models described above, may be explained by either more
efficient removal of enriched gas or a higher average inflow rate of
relatively unenriched gas in the latter. In general, the level and
pattern of abundances for other elliptical galaxies observed with {\it
  Suzaku} are near solar \citep{mat07,taw08,kom09,sat09,hay09} --
similar to what we report here for NGC 4649, indicating that our
simple models may be widely applicable in accounting for the origins
of ISM metals in elliptical galaxies (although with somewhat lower
abundances than typical of X-ray luminous ellipticals, NGC 4649 may be
a somewhat extreme case). A more thorough analysis that compares and
contrasts elliptical galaxy ISM abundances in the context of their
environments and intrinsic properties is beyond the scope of this
paper, but is currently being pursued. In particular, a larger sample
may help in evaluating which SNIa yield set is favored by the X-ray
data (next section). An abundance pattern that displayed a lower level
of dilution would be helpful in this regard. A variation (or lack,
thereof) in the magnitude of dilution from galaxy to galaxy constrains
the nature of the inflow and its relation to elliptical galaxy
evolution (see below).

\subsubsection{The Significance of Ni}

With the effects of SNIa diluted by inflow, it becomes more
challenging to use ISM abundances in ellipticals to constrain SNIa
yields. Even so, predictions for the pattern among the Fe group
elements are sufficiently distinctive that some signature remains, as
we illustrate in Figure 16 for the models in Figure 13b. In
particular, the Ni abundance (both its absolute value, and that
relative to Fe and other elements) is a diagnostic of both the level
of SNIa enrichment and of the SNIa nucleosynthesis prescription (see
also Paper I, PM11). Also plotted in Figure 16 are the Ni/Fe ratios
for fits to RGS and XIS $0-2'$ spectra where the Ni abundance is
allowed to float. At the $\sim$keV temperatures of elliptical
galaxies, these are determined by fitting the shape of the spectrum in
the region where various Fe L (and Ni L) features are generally
blended. As a result, there are concerns that these measurements may
not be robust to assumptions and uncertainties in the plasma code
atomic physics input implemented in spectral fitting packages such as
{\sc Xspec}. As such it will crucial to confirm the high Ni abundance
estimated in this work, Paper I, Ji09, etc. with future high energy
resolution X-ray imaging spectroscopic observations, such as those to
be conducted by {\it Astro-H}, and to improve the accuracy of thermal
plasma modeling of Ni L.

\begin{figure}[ht]
\centering
\includegraphics[scale=0.4,angle=0]{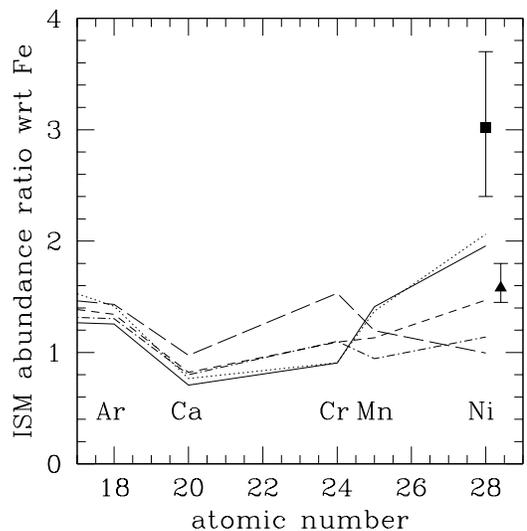}\hfil
\caption{{Same as Figure 13b, focusing on the Fe group
    elements, with Ni/Fe abundance ratios from RGS (solid square) and
    XIS $0-2'$ (solid triangle) spectral fits.}}
\end{figure}

\section{Summary, Implications, and Final Remarks: Results from X-ray
  Spectroscopy of NGC 4649 in the Broader Context of Elliptical Galaxy
  Studies}

The distinctive history of giant elliptical galaxies results in the
creation and maintenance of an extensive ISM, dominated by hot gas,
that contains a fossil record of that very history: evolution
determines ecology, ecology enables archeology, archeology illuminates
evolution. Based on estimates of supersolar stellar $\alpha$-to-Fe
ratios and SNIa rates in excess of $>0.1$ SNU, one naively expects an
ISM abundance pattern in elliptical galaxies that is supersolar across
the board, and increasingly so for elements more proficiently
synthesized in SNIa. In actuality, measured ISM abundances are solar
or subsolar for all elements and do not deviate strongly from solar
ratios, possibly excepting Ni. A simple reduction in the SNIa rate (or
by assuming that SNIa ejecta fail to mix into the hot ISM; Brighenti
\& Mathews 2005) does not resolve this puzzle -- the effective rate
would need be reduced to very low values to explain the modest Fe
abundance {\it level}, begging explanation for why the {\it ratios} do
not then mirror the $\alpha$-element enhanced pattern in the
mass-losing stars.

We bring these issues into sharper focus, and attempt to make progress
on their resolution, through measurement of the abundance pattern in
the elliptical galaxy NGC 4649 derived from deep {\it Suzaku} and {\it
  XMM-Newton} observations. The measurement of abundances in the hot
ISM of elliptical galaxies remains problematic, despite the high
quality of the spectra extracted from these data due to effects of
angular resolution in the case of {\it Suzaku}, background systematics
in the case of {\it XMM-Newton}, and lingering atomic physics
uncertainties (see Appendix A). Nevertheless, by demanding a degree of
cross-mission consistency, and focusing most on conclusions that rely
on relatively model-independent abundance ratios integrated over the
galaxy, we find that robust constraints on elliptical galaxy evolution
can be inferred. Towards this end, we compare the galaxy-wide average
ISM abundance pattern with one-zone steady state solutions to the
equations of chemical evolution that depend on a single parameter that
characterizes the relative contributions of SNIa, stellar mass loss,
and inflow to the ISM metal inventory. We vary this parameter to fit
the data, utilizing the most recent solar standard abundance scale
\citep{agss09} and SNIa yields \citep{mae10}, and reproduce the
observed abundance pattern in NGC 4649, thus lending some insights
into the contributions of various sources of ISM metal enrichment.

Our approach has its drawbacks and limitations, and should be
considered provisional as we expand the data analysis sample and
refine our formalism. By adopting the steady-state limit of equations
(1) and (2), we focus on the relative contributions of stellar mass
loss, SNIa, and inflow/outflow (in the broad sense in which they are
defined here), but gloss over the fact that these are not strictly
co-eval and that the dilution factor refered to as ``inflow'' may in
part correspond to a pre-exisiting group environment that continues to
be reflected in gas phase abundances at large galactic
radii. Encapsulating the relative impact of stellar mass return,
dilution, and SNIa on the pattern of abundance ratios in a single
parameter has the virtue of facilitating accurate model/data
comparison, and emphasizes their interplay and degeneracy -- e.g., the
challenge in distinguishing the combination of low SNIa rate and low
dilution from that of high SNIa rate and high dilution. However, this
complicates the interpretational framework somewhat while, at the same
time, limiting it by leaving these unbroken degeneracies in place. We
are confident that these degeneracies may be broken as the sample, and
our confidence in absolute abundance measurements, grow.

We find that a key ingredient in models that successfully reproduce
the observed pattern is dilution via inflow of subsolar metallicity
extragalactic gas with an (assumed) abundance pattern similar to that
in the stars at an average rate comparable to, or greater than, that
of stellar mass loss. As a result the overall ISM metallicity is
reduced while, at the same time, the effectiveness of Fe group
enhancement from direct SNIa injection is diluted. Although SNIa rates
of 0.1 SNU or more are not required, we now find that they may be
accommodated (and, in fact, that specific rates $<0.02$ SNU are
excluded by limits on the ISM $[\alpha/Fe]$ ratio relative to that in
the stars) provided that a significant fraction of the ISM is of
external origin. Astrophysically reasonable magnitudes of SNIa
enrichment and external dilution may play off against one another, as
shown in Figure 15, to produce subsolar abundances in roughly solar
proportions as observed. In a similar vein, \cite{bmhb09} modeled the
two-dimensional gasdynamics of NGC 4649 and found that inflow of
circumgalactic gas with metallicity comparable to the mean stellar
value could dilute the effects of SNIa enrichment (in their models,
$\theta_{\rm SNIa}/\theta_{\rm MR}=0.84$ using our scale) on the Fe
abundance by a factor of $\sim 3$. We cannot precisely identify the
origin and means of delivery of the extragalactic material. Inflow may
be quasi-steady, or intermittent following the episodes of outflow
that must occur to prevent the gas from accumulating to levels beyond
what is observed \citep{cpm09}.

The reservoir from which gas accretes may be an intracluster,
intragroup, or filamentary intergalactic medium; and may have been
previously ejected from the same galaxy (or other
galaxies). Alternatively, the extragalactic material may originate in
discrete instances of galaxy merging with gas-rich systems. Some
combination of these mechanisms would seem most likely, with their
relative importance depending on galaxy history and environment
\citep{lab11}. The level and pattern of ISM metal enrichment will
reflect this and, with detailed modeling and additional observational
analysis, may prove an important diagnostic of the nature of
post-formation elliptical galaxy interactions.

Thus, the X-ray abundance pattern confirms the complex flow and
enrichment history of NGC 4649 proposed in \cite{bmhb09}, and reflects
the dynamic, continually evolving nature of elliptical galaxies
implied by optical observations \citep{fab07}. The earliest epochs of
elliptical galaxy formation are characterized by the interplay of
major mergers, inflow of primordial gas, vigorous star formation, and
powerful outflows \citep{cpm09,man10,arr10}. Even after the quenching
of star formation has halted the main formation epoch, ellipticals do
not evolve purely in a passive manner in isolation
\citep{tho10}. Evidence of mergers and other interactions, indicate
that ellipticals continue to grow up to the present day
\citep{tra00b,vd08,tal09}, as required to explain the observed
increase in the size of ellipticals and the continual build up of the
red sequence \citep{njo09,vdw09,sha10,rob10,tfd11,dam11,cas11}. In
addition, there are signs of the presence of modest amounts of more
recently formed stars -- often in ellipticals with morphological signs
of interaction \citep{tra00b,kav08,kav09,san09,kav10,veg10}. However,
evolution subsequent to the primary star forming epoch must proceed in
such a way as to be consistent with the old and passively evolving
nature of elliptical galaxy stellar populations inferred from their
individual properties and scaling relations
\citep{pm06,pm08}. Attempts to resolve this apparent paradox, alluded
to in the introduction, often appeal to the process of ``dry'' merging
\citep{tal09,kav10,coo11} that involves very little gas -- and hence
new star formation (as contrasted with the $z>2$ ``wet'' mergers where
stars are efficiently formed). The observed scaling relations and
predominance of old stellar populations are thus preserved. However it
is not clear that mergers of this type occur in sufficient numbers
\citep{hop10}; and, most galaxies interacting with ellipticals are
not, in fact, gas-free \citep{dep10,so10}.

Massive elliptical galaxies are themselves not gas-poor, and the
observed properties of the hot ISM are often neglected in these
considerations. We have shown that the chemical structure of the hot
ISM in ellipticals not only confirms a continual environmental impact,
but demonstrates that the accreted gas is hot or is heated as it is
introduced and interacts with pre-existing hot gas. This implies a
reduction of the star formation efficiency of any cold gas that may be
newly introduced either in the form of mergers with small gas-rich
galaxies or via a smoother accretion process \citep{ker09}.

\lastpagefootnote

The entire one-dimensional ``humidity-based'' classification of
mergers\footnote{in addition to ``wet'' and ``dry'', mergers have also
  been referred to as ``moist'' \citep{san09} and ``damp''
  \citep{for07}} implicitly assumes a one-to-one correspondence
between gas content, dissipation, and star formation that breaks down
in giant ellipticals known to be filled with hot gas. An encounter
involving such a galaxy is never gas-poor, and may involve substantial
dissipation if the companion galaxy is also gas-rich. However, these
mergers may prove to be a hostile setting for new star formation, and
are effectively ``dry.'' Alternatively, the late-time accretion may be
dominated by hot mode accretion \citep{vdV11}. Understanding
elliptical galaxy evolution requires tracking the three ``phases'' of
baryonic matter -- stars, (cold) star-forming gas, and (hot) inert gas
-- and how their mutual exchange of mass, metals, and energy as a
function of age, environment, and type of merger proceeds as the
galaxy evolves subject to internal and external influences
\citep{del10,lu11}.

The rapid build-up of the bulk of the stars that compose present-day
giant elliptical galaxies was quenched by some process that maintained
a galaxy ecology free of significant amounts of cold gas. Suggesed
mechanisms, that also may account for the observe dichotomy in the
characteristics of galaxies, include quenching via gravitational
heating \citep{db08,bd11,nb07,jno09,cat09}, and via AGN feedback (Bell
et al. 2012 and references therein). We see that ISM abudances are a
potential diagnostic of the continuing operation of star formation
quenching.

Our conclusions are driven by the robust analysis results indicating
that ISM abundances are solar or less with ratios that lie between the
$\alpha$-element enriched pattern expected from pure enrichment from
stellar mass loss and the O/Ne/Mg-poor pattern expected from the
combination of stellar mass loss and direct SNIa injection at
$\theta_{\rm SNIa}\sim 1$. As NGC 4649 is a member of the Virgo
Cluster and displays a compact X-ray morphology, the evolution of its
hot ISM (e.g., stripping of the outer regions) may be affected by the
interaction of the galaxy (or a subgroup to which it belongs) with the
ICM. Nevertheless, given the universality of the departure of observed
elliptical galaxy hot ISM abundances from the naive expectations
described above, we suggest that the following main conclusions we
draw for NGC 4649 may be generalized:

\begin{itemize}

\item ISM abundance patterns indicate that metal enrichment from
  stellar mass loss and from direct injection of SNIa, and dilution
  from infall are all significant when averaged over the residence
  time of the ISM. SNIa injection prevents the ISM from reflecting the
  supersolar stellar $[\alpha/Fe]$ ratio, and infall dilutes the
  overall ISM metallicity.

\item An amount of low metallicity gas exceeding that originating
  in stellar mass loss rate must be introduced into the galaxy if one
  is to reconcile the ISM Fe abundance with the standard specific SNIa
  rate of 0.16 SNU.

\item We find further support to the picture of ellipticals as ``open
  ecosystems'' that continually grow, and exchange mass and metals
  with the intergalactic environment and/or other galaxies.

\item The gas that encroaches into or is introduced into the ISM must
  already be hot, or must be heated by mixing with the hot gas and/or
  some other internal mechanism in order to explain why these
  processes are accompanied by little or no star formation.
\end{itemize}

\acknowledgments

We are grateful to Adam Foster, and Randall Smith for their input
regarding atomic physics issues. This work could not have been
completed without support from NASA {\it ADAP}, and {\it Suzaku} and
{\it XMM-Newton} GO programs, and was improved by feedback from a
constructive referee's report.

\appendix

\section{Everything You know is Wrong: Effects of Updating the Atomic
Database}

We initiated analysis of NGC 4649 utilizing the {\em AtomDB} v1.3.2
data base as implemented in {\sc Xspec} version 12.6. Subsequently,
with the availability of the updated {\em AtomDB} v2.0.1 it became
evident that the previous best fit models were no longer statistically
acceptable due to severe differences in Fe L-shell strength
predictions. This is primarily due to differences in the calculated
ionization balance of Fe (Adam Foster, private communication). In this
appendix we conduct simple illustrative numerical experiments to
explore the effect of using the ``wrong'' database to fit spectra, at
CCD energy resolution, of thermal plasmas with $kT\sim 1$ keV. We find
significant discrepancies for $kT<1.3$ keV that -- to the extent that
the update represents a better description of nature -- casts doubt on
the accuracy of previous X-ray results for elliptical galaxies and
other $\sim$keV astrophysical plasmas. The resulting artifacts include
mis-estimates of the plasma temperature and metal abundances, and
false statistical preference for two temperatures.

First we construct simulated spectra, using the NGC 4649 $0-2'$
background and response files. These spectra are normalized to yield
10000 source counts, and are modeled as a solar abundance
\citep{agss09}, single-temperature thermal plasma according to {\em
  AtomDB} v2.0.1. These are then fit with (one- and two-temperature)
models that instead utilize {\em AtomDB} v1.3.2, and 90\% confidence
regions derived by cycling through 1000 such realizations. As a
control, the identical experiment is conducted by repeating with {\em
  AtomDB} v2.0.1 fits. Temperatures of 0.3-2.0 keV are considered.

The ratios, with respect to the input temperature $kT_{in,2.0.1}$, of
the best-fit {\em AtomDB} v1.3.2 temperature $kT_{out,1.3.2}$, are
shown in Figure 17a; $kT_{out,1.3.2}$ versus the best-fit control
temperature $kT_{out,2.0.1}$ in Figure 17b. Figure 18 shows the ratio
of the best-fit Fe abundance in the {\em AtomDB} v1.3.2 fits to those
in the control fits. Other elements show similar behavior, with larger
errorbars. Clearly, temperatures and abundances are underestimated
relative to their ``true'' values, with the effect most prominent
between 0.5 and 1 keV.

In addition, the goodness-of-fit statistic using the ``incorrect''
plasma model is significantly worse relative to the control fits for
$kT<1.3$ keV, as shown in Figure 19a. Given that the resulting
reduced-$\chi^2$ ($\chi^2/\nu$) generally significantly exceeds one,
standard procedure is to add an additional component. We conduct an
experiment along these lines as well for simulation input temperatures
of 0.8, 0.9, 1.0, 1.1 keV and find that, indeed, $\chi^2/\nu$
generally declines (except for $kT_{in,2.0.1}=1.1$) and attains
acceptable magnitudes $\sim 1.1$ when an additional hotter ($\sim
1.2-1.6$) temperature component is considered. This also converts
what had been an abundance {\it underestimate} for one-temperature
models using the ``wrong'' plasma model into an abundance {\it
overestimate}, as demonstrated for Fe in Figure 19b.

\begin{figure}[ht]
\centering
\includegraphics[scale=0.4,angle=0]{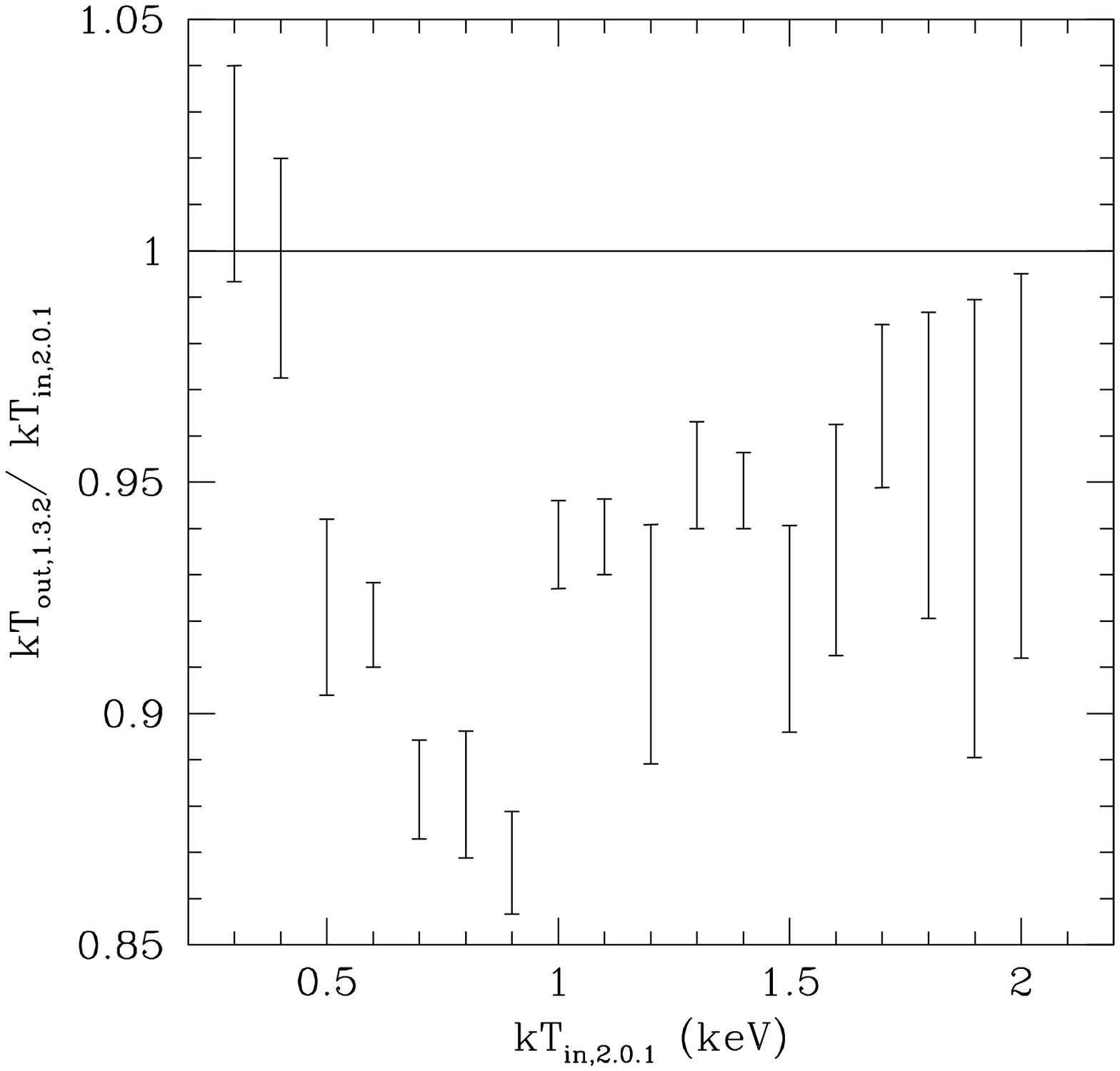}
\includegraphics[scale=0.4,angle=0]{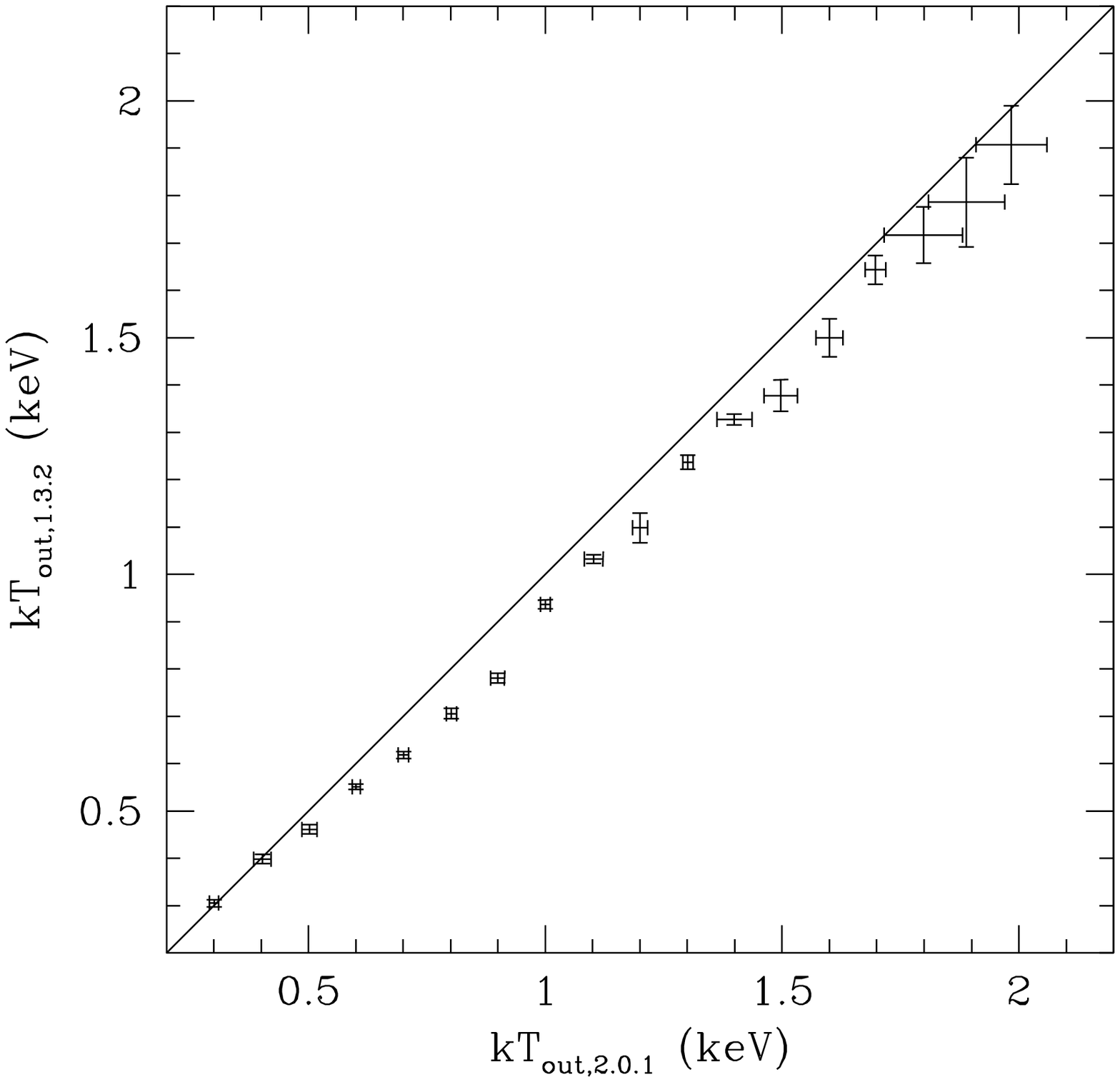}\hfil
\caption{{{\it Left} panel (a): Ratio of the best-fit
    temperature using {\em AtomDB} v1.3.2, to the {\em AtomDB} v2.0.1
    simulation input temperature, versus simulation input
    temperature. {\it Right} panel (b): Best-fit temperature using
    {\em AtomDB} v1.3.2. versus best-fit temperature using {\em
      AtomDB} v2.0.1.}}
\end{figure}

\begin{figure}[ht]
\centering
\includegraphics[scale=0.4,angle=0]{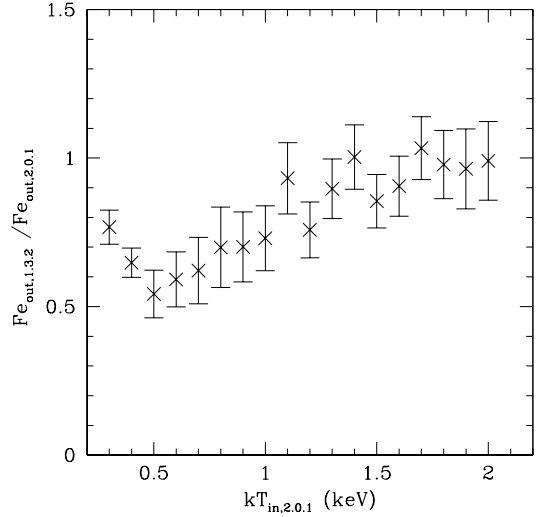}\hfil
\caption{{Ratio of the best-fit Fe abundance using {\em
      AtomDB} v1.3.2, to the best-fit that using {\em AtomDB} v2.0.1,
    versus simulation input temperature.}}
\end{figure}

\begin{figure}[ht]
\centering
\includegraphics[scale=0.4,angle=0]{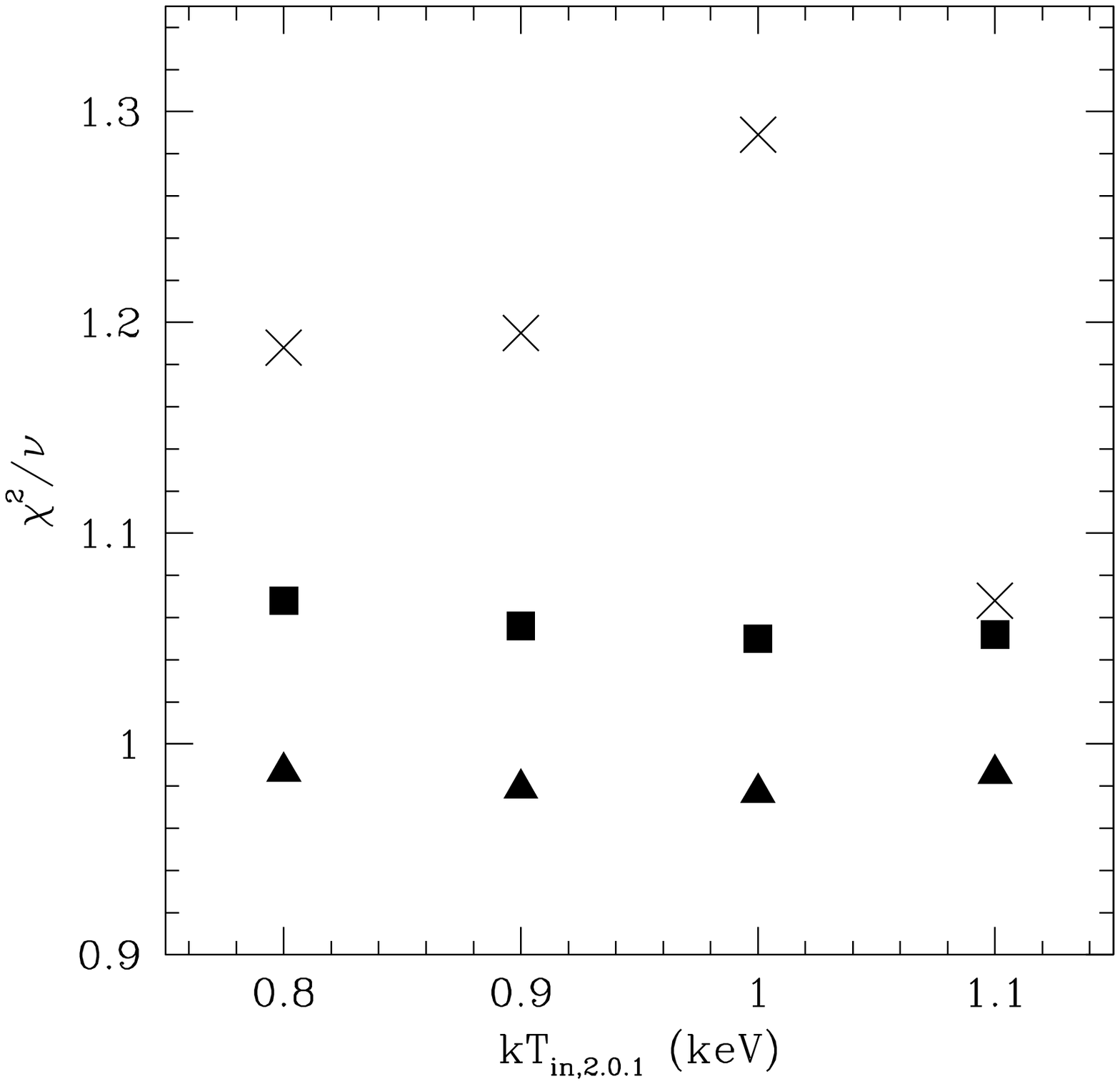}
\includegraphics[scale=0.4,angle=0]{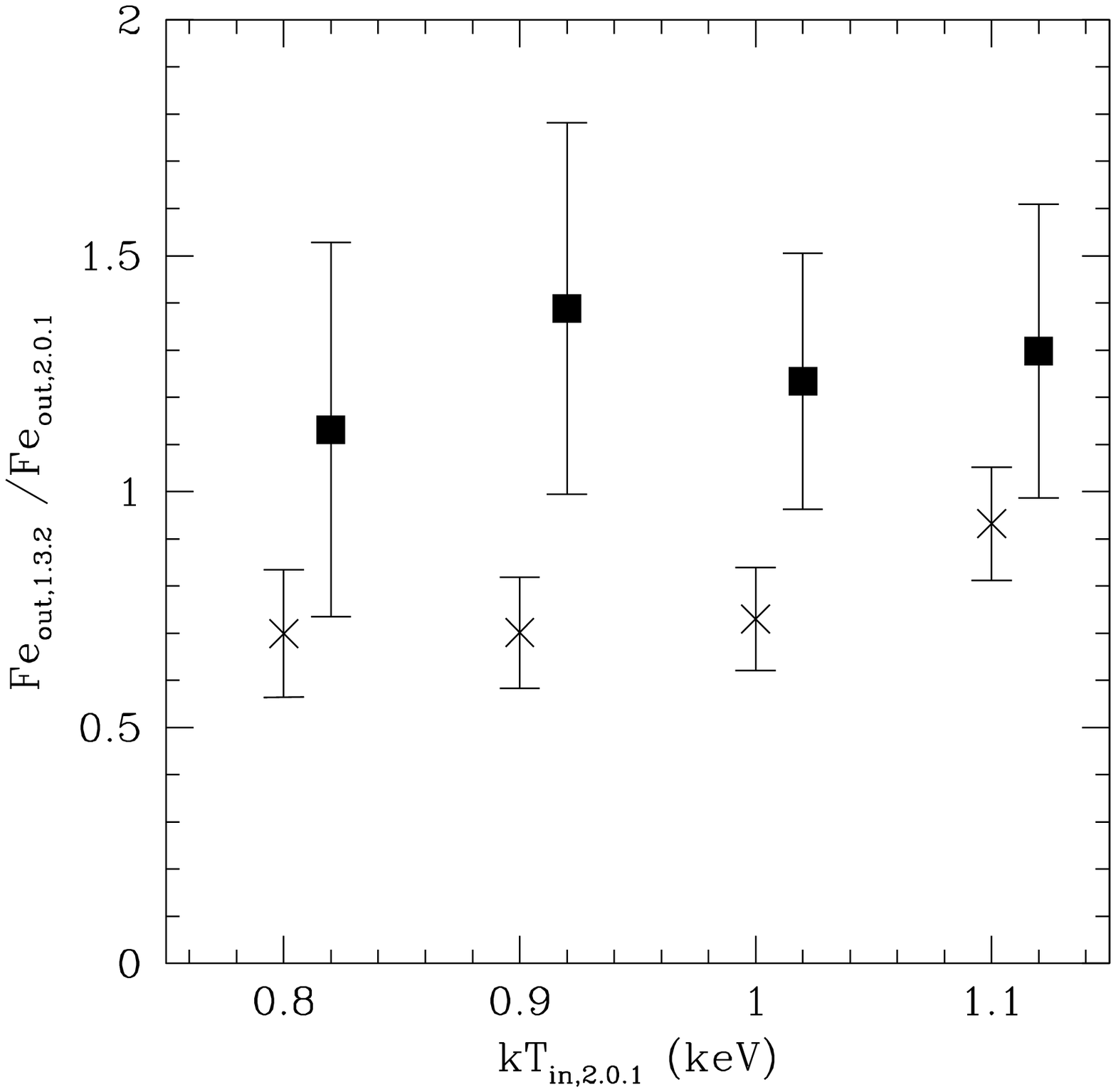}\hfil
\caption{{{\it Left} panel (a): Reduced-$\chi^2$ values
    for fits to simulated {\em AtomDB} v2.0.1 spectra with
    $kT_{in,2.0.1}$ for fits using (the ``correct'') {\em AtomDB}
    v2.0.1 single-temperature models (filled triangles), {\em AtomDB}
    v1.3.2 single-temperature models (crosses), and {\em AtomDB}
    v1.3.2 two-temperature models (filled squares). {\it Right} panel
    (b): Same as Figure 19a for the Fe abundance, showing single-
    (crosses) and two-temperature (filled squares) models.}}
\end{figure}

Based on these simple, illustrative numerical experiences we may
conclude that deficiencies in the earlier versions of {\em AtomDB}
that includes the {\bf apec} thermal plasma model output spectra,
likely resulted in temperature underestimates by as much as
15\%. Metal abundance estimates were underestimated for
single-temperature models, and overestimated for two-temperature
models -- where the latter would have been artificially preferred
based on goodness-of-fit statistics. The effect, while not monotonic
with temperature, is most significant for ``true'' temperatures
$\sim$0.5-1 keV.

Of course, while {\em AtomDB} v2.0.1 is surely more accurate than its
predecessors various, instances of incompleteness and approximation
necessarily remain,\footnote{http://www.atomdb.org/} begging the
question of the accuracy of model parameters obtained in current
spectral fits.

When models of astrophysical plasma spectra are formally unacceptable,
this may be attributed to complexities in the physical conditions or
to inaccuracies in the plasma model. Distinguishing these explanations
is facilitated by improvements in the quality of atomic databases and
X-ray spectra. At the same time, confrontation between the models and
data can point the way to improvements in the accuracy of both our
understanding of the plasma physical conditions and the key remaining
deficiencies in the atomic database and its application.

%==============================================================================

\clearpage

\end{document}